# On Magneticons and some related matters[1]

David Fryberger

*SLAC National Accelerator Laboratory*
*2575 Sand Hill Road, MS-62, Menlo Park, CA USA 94025*
Email address: Fryberger@SLAC.stanford.edu

The name "magneticon" in this paper refers to a magnetically charged spin ½ particle predicted by incorporating a symmetry of classical electromagnetism, called dyality symmetry, into a certain model for the structure of point-like fermions. (Actually, it is anticipated that there would be a full spectrum, both hadronic and leptonic, of such magnetically charged particles.)  The lightest of these magneticons is anticipated to be leptonic in nature and predicted to have the same magnitude of electromagnetic charge (in Gaussian units) as the electron, except that it is magnetic.  Accompanying this spectrum of magnetic fermions, it is suggested that there may also be a second, or magnetic, photon.  After a brief introduction, the pair production cross section of magneticons by electron-positron annihilation is derived using a lowest order quantum perturbation approximation suggested by a two-potential Lagrangian form for classical electromagnetism, symmetrized through the use of space-time algebra to include magnetic charge and currents.  A discussion of how these ideas might be included in other quantum interactions involving magnetic charges and the magnetic photon is undertaken.  These interactions include electron-magneticon scattering and magneticon vacuum polarization loops.  Possibilities for the observation of these magnetic particles in past experiments, as well as future experiments, are explored, and some predictions are made.

## I. INTRODUCTION

In this paper, the name "magneticon" refers to a magnetically charged spin ½ particle that is predicted to have the same magnitude of electromagnetic charge as the electron except that it is magnetic, i. e., either north or south.[1]  More generally, a magneticon would be a magnetically charged fermion belonging to a full spectrum of "magnetic" fermions predicted to exist as counterparts (on a one-to-one basis) to those of the Standard Model (SM).  That is, if this dyality[2] symmetry of generalized electromagnetism  is realized in Nature, then, one also expects magnetic counterparts to all of the fundamental SM fermions, e. g., magnetic muons, magnetic neutrinos, magnetic protons, magnetic neutrons, etc. (as well as their antiparticles). The lightest of these (charged) magneticons would presumably be leptonic, and hence would be a magnetic electron.[3]  (At times, we shall use the word magneticon as a specific reference to the magnetic electron.)  Expanding upon this idea, the prediction of a second (or "magnetic") set of particles, counterparts on a one-to one basis to those of the SM, is effected by applying the operator $\exp(\gamma_5 \, \Theta_d)$, where the dyality angle $\Theta_d = \pm\pi/2$, to the spectrum of SM fermions as described by "A Model for the Structure of Point-Like Fermions" [3].  Extending this prediction to the bosons, then, one would also predict the existence of a magnetic photon[4] and magnetic Zs and Ws, as well as magnetic Higgs particles. Except for the magnetic photon, which by gauge invariance should be massless,[5] the presumed reason for the non-observation of these magnetic counterparts would be that they are too massive. (Compounding difficulties, discussed below, are the specific experimental triggering and tracking algorithms.) To be consistent with this viewpoint, it would be appropriate, then, to classify the present SM particles (including the photon) as "electric," and these newly predicted particles (including a second photon) as "magnetic."[6]

---

[1] This material is based upon work supported by the U. S. Department of Energy, Office of Science, under Contract No. DE-AC02-76SF00515.

## II. SOME BACKGROUND

Twenty some years ago, a classical Lagrangian formalism was given from which both the symmetrized set of Maxwell's equations and the equations of motion for both electrically and magnetically charged particles can be derived[7] [6]. This is an interesting result because there are a number of earlier papers that claim such a derivation either is not possible, or assert that certain restrictions on the behavior of the electric and magnetic charges are required, e. g., Refs. [7 - 11]. The analysis in Ref. [6] used two potentials, as put forth by Cabibbo and Ferrari [4] and others [2, 12], and employed space-time algebra,[8] the Clifford algebra appropriate to four-dimensional space-time [14]. Gaussian units (See, e. g., Ref. [15]) are used in Ref [6]; they are particularly convenient because of the symmetrical treatment of electric and magnetic quantities. For the convenience of the reader, we give below a brief overview of the underlying Clifford algebraic basis for the derivation of the $e^+e^-$ annihilation cross section for magneticon pair production, as well as other quantum calculations.

Four linearly independent vectors $\gamma_\mu$ ($\mu = 0,1,2,3$) are used as a basis set for space-time algebra. The (Clifford) products of these vectors yield 16 linearly independent quantities, which partition into scalar, vector, tensor, axial vector (or pseudovector), and pseudoscalar objects, in complete analogy to the bilinear forms which can be constructed used using solutions to the Dirac equation.

The pseudoscalar of space-time algebra, $\gamma_5$, is defined by

$$\gamma_5 = \gamma_0 \gamma_1 \gamma_2 \gamma_3. \tag{1}$$

With this definition $(\gamma_5)^2 = -\mathbf{1}_4$, where $\mathbf{1}_4$ is the 4×4 unit matrix.[9] Of course, $\gamma_5 \gamma_\mu = -\gamma_\mu \gamma_5$ still obtains. Thus, in space-time algebra $\gamma_5$ plays a role analogous (in some versions of electromagnetic theory) to that of the imaginary quantity $i = \sqrt{-1}$. [We note here that the $\gamma_5$ used in Eqs. (2-4) is that of space-time algebra as defined by Eq. (1); for convenience, the subsequent QED and QEMD calculations in this paper will use the $\gamma_5$ as defined by Eq. (16).]

It was shown in Ref. [6] that all of the relevant equations of the generalized electromagnetism have a continuous symmetry described by an arbitrary angle.[10] This symmetry, which we call dyality symmetry, is manifest by an invariance of form when all terms of an equation are multiplied by the factor $\exp(\gamma_5 \Theta_d)$, where the dyality angle $\Theta_d$ can be arbitrarily specified.[11] In this way, the quantity $\gamma_5 \Theta_d$ effects a rotation in the electromagnetic plane. In particular, electric and magnetic quantities are exchanged when $\Theta_d = \pm \pi/2$, converting a theory of electromagnetism into a theory of "magnetoelectricity," and vice versa.

The (classical) electromagnetic interaction term of the Lagrangian mentioned above is the Clifford product of a generalized current density vector

$$\mathcal{J} = j - \gamma_5 k \tag{2}$$

times a generalized (four) potential

$$\mathcal{A} = A - \gamma_5 M, \tag{3}$$

where $A$ is the usual vector potential associated with the electric current density vector $j$, and $M$ is the magnetic vector potential associated with the magnetic current density vector $k$. The space-time algebraic expansions of these quantities are $j = j^\mu \gamma_\mu$, $A = A^\mu \gamma_\mu$, etc., where $j^\mu$, $A^\mu$, etc. are the usual quantities in tensor analysis.

When written out, this generalized interaction term (note the several minus signs) is

$$-\mathcal{J}\mathcal{A} = -(j - \gamma_5 k)(A - \gamma_5 M) = -(jA - j\gamma_5 M - \gamma_5 kA + \gamma_5 k \gamma_5 M). \tag{4}$$

The usual interaction term, $-jA$, describes the interaction of an electric current with the usual vector potential.[12] The term $-\gamma_5 k \gamma_5 M = -kM$ is the analogous interaction of a magnetic current with a magnetically generated (magnetic) vector potential. This would be the interaction to be used in what might be called QMD, the quantum theory of an exclusive magnetic world. The cross terms, $j\gamma_5 M + \gamma_5 kA$, describe the forces of magnetically generated fields on electric currents, and vice versa. These terms lead to the cross term forces in a generalized Lorentz force equation. And it is these terms that we seek to incorporate into an extension of QED, which we propose to call QEMD.[13] That is, we presently exist in an electric world, but wish to incorporate in a consistent way into QED the possible existence of magneticons and of a second, or magnetic, photon into this electric world, hence the name QEMD.

In these QEMD analyses, we look to Ref. [6] for further guidance. In this regard, in Ref. [6] we see that classical Maxwell theory clearly shows that electric charge and current, i. e., $j$, is the source for the vector potential $A$ as well as its associated electromagnetic field tensor $F_{\mu\nu}$. Therefore, we take the view that the QED vertices that we would identify as Maxwell source (MS) terms can be thought of as photon emission vertices. These emitted fields are, of course, to be associated with the (electric) photon of quantum theory. When the generalized classical Maxwell theory is formulated to include magnetic charge and current, we see that the analogous magnetic quantity $k$ is the "Maxwell" source for the magnetic vector potential $M$ and its associated magnetoelectric field tensor $G_{\mu\nu}$. These magnetically sourced field quantities, then, are to be associated with a second, or magnetic, photon (via MS vertices). Again, we view these magnetic MS vertices in QEMD (or QMD) as magnetic photon emission vertices.

Continuing our search for guidance in the generalized classical Maxwell theory of Ref. [6], we observe that it is these electromagnetic fields (and their magnetic counterparts) that exert forces on electric (and magnetic) charges and currents via the Lorentz force equation for which the interaction terms are given in Eq. (4). In QEMD, then, one can think of the vertices associated with the Lorentz force cross (LFc) terms as photon absorption vertices. In order to properly bring into quantum mechanics the magnetic charges, currents and fields, it is important to distinguish which vertices we are representing in the quantum theory: the analogues to 1) the MS term or to 2) the Lorentz force term.

It is also important to observe that in the generalized theory that we are presenting, the electric and magnetic photons and their fields are physically distinct; that is, the electric fields associated with the electric charges and currents are distinct from those electric fields associated with magnetic charges and currents. The same statement also holds true for the magnetic fields associated with the two types of photons. Ref. [6] shows that this distinction between electromagnetism and magnetoelectricity is found in the intrinsic field parities.[14] This is the reason that we have introduced a separate magnetoelectric tensor $G_{\mu\nu}$.[15] Acknowledging that Eq. (4) implies that we have two photons means that there are two possible ways that electric and magnetic particles can interact in QEMD. One can represent these two possibilities by drawing two different Feynman diagrams, as in Fig. 1, both of which represent a possible electron-magneticon scattering interaction. If it is appropriate to view these two diagrams as two different topologies, then, as is shown in Appendix A, the e-m scattering cross section will be roughly doubled: the two topologies would be a direct consequence of there being two photons.

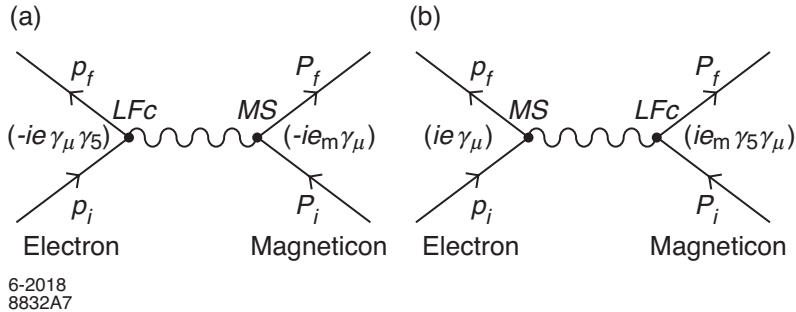

Fig 1. (a) Feynman diagram for e-magneticon scattering assuming that the magneticon is the Maxwell source term (labelled by MS), which then acts on the electron via a Lorentz force cross term (labelled by LFc). (b) Feynman diagram for e-magneticon scattering assuming that the electron is the Maxwell source term (labelled by MS), which then acts on the magneticon via the Lorentz force cross term (labelled by LFc). For both (a) and (b), the QEMD vertex factors are indicated in parentheses. Note that our choice of $e$ to be the positron charge will lead to some sign differences with other formulations (e. g. Ref. [21]). The initial (final) state is at the bottom (top) of the diagram.

In a prior paper [20], following as a template an analysis of electron-Coulomb scattering [21], the above equations and relationships were used to formulate a naïve[16] quantum mechanical calculation of electron-monopole scattering, which was then shown to be consistent with the well-studied classical calculations.[17] Looking back at this paper, the magnetic potential, which derives from generalized Maxwell's equations, is made explicit in writing the magnetic "Coulomb" potential as
4

$$M_0 = \frac{g}{4\pi |\boldsymbol{x}|} \quad \text{with} \quad \boldsymbol{M} = 0, \tag{5}$$

where $g$ is the magnetic charge of the monopole (We use bold font for 3-vectors, and natural units for which $\hbar = c = 1$.) In that analysis, the electron (current) is scattered (via a LFc term) by a magnetic "Coulomb" potential acting as a magnetic source (a magnetic MS term). (See Fig. 1a) Following this logic, the transition matrix element for this $t$-channel process[18] was written as

$$S_{\text{fi}}^{\text{eg}} = -ie \int d^4 x \, \bar{\psi}_{\text{f}} M_\mu \gamma^\mu \gamma_5 \psi_{\text{i}}, \tag{6}$$

where, as in Ref. [21], $e$ is the charge of the electron ($e < 0$) and $\psi_{\text{i}}$ ($\bar{\psi}_{\text{f}}$) represents the initial (final) electron wavefunction.[19] Consistent with the analogous ($j\gamma_5 M$) LFc term in Eq. (4), the $\gamma^\mu$ precedes the $\gamma_5$.[20] For later use, we record Eq. (14) of Ref. [20], after averaging over the initial spins and summing over the final spins (i. e., taking the trace), for e-g ("Coulomb") scattering in this model, which is:

$$\frac{d\bar{\sigma}_{\text{eg}}}{d\Omega} = \frac{g^2 \alpha}{8\pi |\boldsymbol{q}|^4}(8E_{\text{i}}E_{\text{f}} - 4p_{\text{i}} \cdot p_{\text{f}} - 4m_{\text{e}}^2) = \frac{\alpha_{\text{m}}\alpha}{2|\boldsymbol{q}|^4}(8E_{\text{i}}E_{\text{f}} - 4p_{\text{i}} \cdot p_{\text{f}} - 4m_{\text{e}}^2), \tag{7}$$

where $\boldsymbol{q}$ is the three-vector momentum transfer, and we have used Eq. (23), below, to render this result applicable to the magneticon study herein. Here, $\alpha = e^2/4\pi$, and $\alpha_{\text{m}} = g^2/4\pi$. Note that the $m_{\text{e}}^2$ term is of opposite sign to that of the analogous result for electron Coulomb scattering [21, Eq. (7.21)]. Such sign reversals of mass terms are characteristic of the LFc terms in this model. The final result for e-g scattering in the lab frame (with the magnetic charge, of infinite mass, at rest) is

$$\frac{d\bar{\sigma}_{\text{eg}}}{d\Omega} = \frac{\alpha_{\text{m}}\alpha \cos^2(\theta/2)}{4\boldsymbol{p}^2 \sin^4(\theta/2)}, \tag{8}$$

where $\boldsymbol{p}$ is the initial electron 3-momentum and $\theta$ is the electron scattering angle. This is the same result as found in Appendix A, Eq. (A25), which was based on a more general analysis (finite mass proton). Comparison of Eq. (8) to that for electron Coulomb scattering, which is [21]:[21]

$$\frac{d\bar{\sigma}_{\text{eQ}}}{d\Omega} = \frac{\alpha^2 [1 - \beta^2 \sin^2(\theta/2)]}{4\boldsymbol{p}^2 \beta^2 \sin^4(\theta/2)}, \tag{9}$$

reveals that the factors of $\beta^2$ have, in effect, been replaced by 1. (In this instance $\beta = v/c$, and refers to the incident electron.) This result is anticipated by recalling that the classical Lorentz force (cross) term contains the product $\boldsymbol{v} \times \boldsymbol{B}$. The question of a two photon cross section is further explored in Appendices A and B. [Except for Eq. (6), we shall use $e$ to represent the positron charge.][22]

### III. THE MAGNETICON PAIR PRODUCTION CROSS SECTION

Another quantum mechanical perturbation calculation of interest here has an initial electron-positron state annihilate to a virtual photon via the $jA$ interaction, with the virtual photon subsequently creating a magneticon pair via a $\gamma_5 kA$ LFc term interaction. (See Fig. 2b.) Our QEMD interpretation of this s-channel Feynman diagram, then, is that the initial vertex (the $jA$ interaction) is in the form of what we are calling a Maxwell source term (or MS term), producing a (virtual electric) photon (the vector potential $A$), which then interacts with the (magneticon) current $k$ (flowing in the vacuum), which is promoted to a pair of real final state magneticons. This latter interaction vertex is categorized as a Lorentz force cross term (or LFc term). To be consistent with the above generalized electromagnetism analysis, we include in the quantum interaction description of this LFc term a $\gamma_5$, indicating an axial vector interaction, as shown in Fig. 2b.

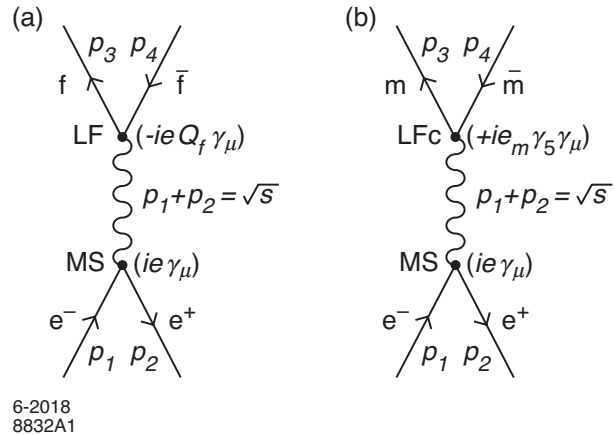

6-2018
8832A1

Fig. 2. Feynman diagrams (CM frame in momentum space) for $e^+e^-$ annihilation: (a) to a heavy fermion-antifermion pair of charge $\pm Q_f e$; (b) to a magneticon pair of charge $\overset{N}{\underset{S}{}} e_m$. In both diagrams, the propagating photon is electric, produced at the initial annihilation vertex by a Maxwell source term associated with an electric current. Hence, this vertex is labelled MS. The relevant QED(a) and QEMD(b) vertex factors are given in parentheses. The use of the $\gamma_5$ at the magneticon pair production vertex in Fig. 2b derives from the fact that this vertex is, as described in the text, a Lorentz force cross term vertex; hence this vertex is labelled LFc. The initial (final) state is at the bottom (top) of the diagram. There is further discussion of these diagrams in Appendix A.

At this point, following our precepts derived from the classical generalized E&M, we argue that the magnetic current in the vacuum is not free and hence cannot be a source[23] for a magnetic photon that clearly would have to propagate backward in time to connect to the (globally) prior $e^+e^-$ annihilation vertex. Thus, in this QEMD s-channel interaction, we have only one (active) topology. We acknowledge that this view appears[24] to introduce a violation of time reversal invariance into QEMD physics.[25] Similarly, the fully reversed interaction consisting of $m\bar{m} \to {}_m\gamma^* \to e^+e^-$ would also be a one-topology interaction, and its evaluation would go through in the same manner as would $e^+e^- \to {}_e\gamma^* \to m\bar{m}$. ($\gamma^*$ indicates a virtual photon.)

Thus, reviewing the above, we now have a clear argument why we have not yet seen any magnetic photons: we live in an electric world with (essentially) no magnetic charges or currents. That is, there are (essentially) no free magnetic Maxwell sources,[26] and hence no magnetic photons. How is it that the significance of the distinction between the MS vertex and the LF vertex is not made in QED? In this regard, it is of significant interest to observe, for example, that Bjorken and Drell [21, Sec. 7.4] calculate the scattering of an electron by a proton using the MS term at the proton vertex and the LF term at the electron vertex, obtaining an expression for the *S*-matrix element $S_{fi}$. They then calculate $S_{fi}$ with the opposite assignment of vertices, obtaining the same result for $S_{fi}$. They therefore conclude that it doesn't matter which way we look at this scattering interaction, and that we only need a single Feynman diagram to represent this electron-proton[27] interaction, no matter which way one labels the vertices. In fact, one does not even need to label the vertices. By this reasoning, this physics distinction, which is present in classical Maxwell theory, is superfluous in QED, and does not find its way into the well-known Feynman rules of Quantum Mechanics. Another way to characterize this situation is to observe that this physics distinction does not lead to topologically distinct diagrams in e-p scattering, but it (evidently) does in e-m scattering. Following these ideas, the introduction of magneticons and magnetic photons into other QEMD interactions is further explored below.

To proceed, we will need to have an understanding of magneticon pair production. To furnish a specific framework in which to derive magneticon pair production, we first review heavy fermion pair production, which is a well-studied *s*-channel process. For this exercise, we follow the analysis (and notation: natural units, used by Ref. [21], for which $\hbar = c = 1$) by Renton [26], in which an initial $e^+ e^-$ state annihilates into a virtual (electric) photon, which subsequently produces the final state $f\bar{f}$ pair. The relevant Feynman diagram is shown in Fig. 2a.[28] Following the above discussion, the initial vertex here is what we would now call a MS vertex, and the final vertex would be called a LF vertex, but as we have mentioned above, in the Feynman rules of QED, one doesn't need to make this distinction.

The heavy fermion-antifermion pair in the final state is assigned the charge $\pm Q_f e$. (Renton defines $e > 0$.) For the diagram in Fig. 2a, Renton writes for the Lorentz invariant amplitude

$$\mathcal{M}_{fi} = -ie^2 Q_f (\bar{v}_2 \gamma^\mu u_1 \bar{u}_3 \gamma_\mu v_4)/s, \tag{10}$$

where *s* is the energy squared in the center of momentum (CM) frame. The subscripts on the four-component spinors (*u* and *v*) identify the appropriate initial and final state fermion legs. Squaring the magnitude of $\mathcal{M}_{fi}$ yields

$$|\mathcal{M}_{fi}|^2 = \frac{e^4 Q_f^2}{s^2} [(\bar{v}_2 \gamma^\mu u_1 \bar{u}_1 \gamma^\nu v_2)(\bar{u}_3 \gamma_\mu v_4 \bar{v}_4 \gamma_\nu u_3)]. \tag{11}$$

Introducing traces (and summing over final state spins), Eq. (11) becomes

$$|\mathcal{M}_{fi}|^2 = \frac{e^4 Q_f^2}{s^2} \text{Tr}[\not{p}_2 \gamma^\mu \not{p}_1 \gamma^\nu] \text{Tr}[(\not{p}_3 + m_f)\gamma_\mu (\not{p}_4 - m_f)\gamma_\nu], \tag{12}$$

where $m_f$ is the heavy fermion mass. (The electron mass has been neglected.) Including a factor ¼ to average over initial state spins and evaluating the traces yield

$$|\mathcal{M}_{\text{fi}}|^2 = \frac{8e^4 Q_{\text{f}}^2}{s^2}[(p_1 \cdot p_4)(p_2 \cdot p_3)+(p_1 \cdot p_3)(p_2 \cdot p_4)+m_{\text{f}}^2(p_1 \cdot p_2)]. \qquad (13)$$

Going to the CM frame, choosing the z-axis in the direction of the incoming e⁻, and defining $\theta$ as the angle between this direction and that of the outgoing f, one has for the differential cross section

$$\frac{d\sigma_{\text{f}\bar{\text{f}}}}{d\Omega} = \frac{Q_{\text{f}}^2 \alpha^2 \beta}{4s}\left(1+\beta^2 \cos^2\theta + \frac{m_{\text{f}}^2}{E^2}\right), \qquad (14)$$

where $E$ is the beam energy, and the Lorentz invariant two-body phase space factor

$$\mathscr{P}_2 = \frac{p_3 d\Omega}{16\pi^2 W} \qquad (15)$$

has been used. $W = 2E$ is the CM frame energy, and $\beta$ represents the relativistic velocity factor, $p_{\text{f}}/E_{\text{f}}$, of the produced heavy fermions. In the relativistic limit, $\beta \to 1$, and Eq. (14) yields the well-known final state angular distribution $(1+\cos^2\theta)$. In the natural units of Renton's notation, the dimensionless fine structure constant $\alpha = e^2/4\pi \cong 1/137$. ($\alpha = e^2/\hbar c$ in Gaussian units.)

We now turn to magneticon pair production, as depicted in Fig. 2b. As before, we have an initial e⁺ e⁻ pair state annihilating into a virtual (electric) photon. Consistent with the above discussion (and with the f$\bar{\text{f}}$ diagram of Fig. 2a), we view this vertex to be an MS vertex. Actually using this label does not affect the mathematics or the physics of this vertex; we still have an electric current producing the electric photon (described by a vector potential $A_\mu$) with the usual vertex factor $-iq\gamma_\mu$. [Recall we have defined $e$ (= $-q$ in this instance) as the positron charge.] This electric photon, then produces at the final vertex, which is in the form of a QEMD LFc term and hence introduces the factor $+ie_m\gamma_5\gamma_\mu$ (where $e_m$ is the magneticon charge), the final state m$\bar{\text{m}}$ pair. (A magnetic current $k$ in the vacuum is caused to materialize into a real magneticon final state N-S pair.) We use the ordering $\gamma_5\gamma_\mu$ to be consistent with the $+\gamma_5 kA$ ordering for this contribution to the generalized Clifford algebraic formulation, above.

Again, we follow the lead afforded by Ref. [6],[29] but in this instance we have introduced only the Lorentz force cross term,[30] but not the second, or magnetic, photon. To briefly summarize, we argue that an electron (or proton, or any other charged SM particle of charge $q$) can be a source for (i. e., emit) only electric photons at an MS vertex, which vertex then carries the usual QED factor $-iq\gamma_\mu$, while magneticons can be a source for (i., e., emit) only magnetic photons (also via an MS term), but with the quantum magnetodynamics (QMD, or QMED – which would be the quantum theory of a magnetic world with electric charges and photons adjoined) vertex factor $-ie_m\gamma_\mu$. In completing this analogy of the generalized Clifford formulation to QEMD (and to QMED), we see that what we are calling the LF interaction terms[31] (both straight and cross) are to be associated with what we generally think of as photon absorption. We see that this LF term (photon absorption at a vertex) is consistently described whether we are looking at the straight terms of QED or QMD or the cross terms of QEMD or QMED. Looking at the Clifford interaction terms given above [Eq. (4)], we argue that it is these cross term absorption vertices which are to carry a $\gamma_5$ as part of the absorption vertex term.

At this stage, we introduce the standard quantum definition of $\gamma_5$, bringing this pseudoscalar quantity in the usual way into our quantum mechanical analysis. That is [21, p. 282]:

$$\gamma_5 = i\gamma_0\gamma_1\gamma_2\gamma_3 = \gamma^5 \text{, with } (\gamma_5)^2 = 1. \tag{16}$$

The form of the cross terms in the Clifford analysis [i. e., Eq. (4)] will be used to determine the specifics of the quantum mechanical cross terms, i. e., where to put in the $\gamma_5$, as we illustrate in Eq. (17).

With the above background, we continue our investigation of magneticon pair production. Thus, for magneticon pair production, Eq. (10) becomes

$$\mathcal{M}_{\text{fi}}^{(m\bar{m})} = + iee_m(\bar{v}_2\gamma^\mu u_1 \bar{u}_3\gamma_5\gamma_\mu v_4)/s. \tag{17}$$

Squaring the magnitude of $\mathcal{M}_{\text{fi}}^{(m\bar{m})}$ yields

$$\left|\mathcal{M}_{\text{fi}}^{(m\bar{m})}\right|^2 = \frac{e^2 e_m^2}{s^2}[(\bar{v}_2\gamma^\mu u_1\bar{u}_1\gamma^\nu v_2)(\bar{u}_3\gamma_5\gamma_\mu v_4\bar{v}_4\gamma_5\gamma_\nu u_3)], \tag{18}$$

which then becomes

$$\left|\mathcal{M}_{\text{fi}}^{(m\bar{m})}\right|^2 = \frac{e^2 e_m^2}{s^2}\text{Tr}[\not{p}_2\gamma^\mu \not{p}_1\gamma^\nu]\text{Tr}[(\not{p}_3 + m_m)\gamma_5\gamma_\mu(\not{p}_4 - m_m)\gamma_5\gamma_\nu], \tag{19}$$

where $m_m$ is the magneticon mass. (Again, the electron mass has been neglected.) Now, eliminating the $\gamma_5$ factors, and evaluating the traces, we have

$$\left|\mathcal{M}_{\text{fi}}^{(m\bar{m})}\right|^2 = \frac{8e^2 e_m^2}{s^2}[(p_1\cdot p_4)(p_2\cdot p_3)+(p_1\cdot p_3)(p_2\cdot p_4) - m_m^2(p_1\cdot p_2)], \tag{20}$$

where we see that the inclusion of the $\gamma_5$ factors has reversed the sign of the $m_m^2$ term. (A similar result was found in Ref. [20].) We note that while there is only one $\gamma_5$ factor in Eq. (17), a second one is automatically introduced when, in order to obtain the cross section, one squares the amplitude, yielding Eq. (18). Eq. (20), then, yields the differential cross section for magneticon pair production:

$$\frac{d\sigma_{m\bar{m}}}{d\Omega} = \frac{\alpha\alpha_m\beta}{4s}\left(1+\beta^2\cos^2\theta - \frac{m_m^2}{E^2}\right), \tag{21}$$

where $\alpha_m = e_m^2/4\pi$, or in Gaussian units $e_m^2/\hbar c$. Using the relativistic relationship $m^2 = E^2 - p^2$, we can evaluate the mass term in Eq. (21) to get a final result:

$$\frac{d\sigma_{m\bar{m}}}{d\Omega} = \frac{\alpha\alpha_m \beta^3}{4s}\left(1+\cos^2\theta\right). \tag{22}$$

Eq. (22) should be compared to Eq. (14), above. (Keep in mind that $Q_f^2 = 1$.) At energies for which the mass of the produced pairs becomes negligible, these two distributions become identical. However, differences are to be expected near threshold. The threshold behavior of the magneticon pair cross section is softer than standard fermion pair production (an additional factor of $\beta^2$, as one might anticipate by analogy to the $v \times B$ term in the classical Lorentz force equation) and that above threshold the final state angular distribution (not including radiative and other corrections) is always $(1 + \cos^2\theta)$. These two aspects of the functional form are therefore predicted, and should one be able to detect magneticon pair production, they should be observable. Here, we reiterate the argument (see endnote 18) that for this *s*-channel interaction there is no second, or magnetic, photon associated with magneticon pair production.[32] This is in contrast to the *t*-channel e-m scattering described above and further discussed in Appendix A, for which both the electron and the magneticon exist as real particles in both the initial state and the final state, and which, therefore, enables support for both electric and magnetic Maxwell source vertices. Hence, we suggest that for e-m scattering in the 2γ formulation (see below), we have two topologically distinct Feynman diagrams.[33]

We argue, then, that for magneticon pair production, the initial vertex represents a Maxwell source (MS) term and the final vertex represents a Lorentz force cross (LFc) term. This statement is true for both QED and for QEMD. In the case of QED pair production, which is an *s*-channel process, the label assignment doesn't matter.[34] However, for QEMD pair production, we argue that it does matter, and, as we have seen, it affects the result of the calculation. The initial vertex, which is connected to the initial state $e^+e^-$ leptons, has to remain the MS vertex, and the final vertex remains the Lorentz force cross (LFc) vertex, which, as explained above, contains a $\gamma_5$ factor. As we argued in the text, to invert, internally in the calculation, the labelling of these vertices is not appropriate. Thus, we still have one Feynman diagram with the resultant calculation as given by Eq. (22). As a comment, we note that in the configuration space[35] Fourier integrals, for these pair production diagrams, there are regions in which the two vertices are in inverted time order. But we assert that the arguments reconciling this fact with the notion that the annihilation vertex can generally be viewed as prior to the production vertex are as valid in QEMD as it is already accepted that they are in QED. Thus, the QEMD calculation goes through as indicated without internally shifting the location of the $\gamma_5$.[36] Of course, as mentioned above, this point of view needs experimental confirmation.

For the present, we assume that the (manifestly) broken dyality symmetry[37] is not in the magneticon charges, but in the magneticon masses. Thus, invoking dyality symmetry between electromagnetism and magnetoelectricity enables the prediction that

$$|e_m| = |e|, \text{ i. e., } \alpha_m = \alpha .^{38} \qquad (23)$$

This means that as $\beta \to 1$, it is predicted that magneticon pair production would lead to an additional unit of $R$ for each species of magneticon (where we define $R$ to include magneticons) in the total $e^+e^-$ cross section. Should magneticons be observed, it is of importance that these questions of detail be studied experimentally. (One recalls the early questions about vacuum polarization loops as called for in QED with positrons [27]. Experimental confirmation in that case was crucial.)

**IV. EXPERIMENTAL DATA**

A. Some remarks

This section examines several categories of experimental data that have possibilities for revealing the presence of magneticons. Taking inspiration from the original papers of Dirac [1], which show that there may exist magnetic monopoles of charge $g = n\hbar c/(2e)$, where $n = 1, 2, \ldots$, monopole searches have generally used high ionization as a criterion for the event trigger and particle identification. Since this present work considers magneticons with magnetic charge $g = e$ (i. e., not the monopole as conceived by Dirac), it is worthwhile to review and expand upon the expected $dE/dx$ associated with magnetic charge passing through matter. This discussion, found in Appendix B, is based upon the calculated electron-magneticon cross sections found in Appendix A. We first derive a formula for the $dE/dx$ of magneticons that (essentially) agrees with that of Ahlen [22]. We call this the one-photon (1γ) formulation. In addition, we find that if there would exist two photons, one electric and one magnetic, as suggested by the dyality invariance of generalized electromagnetism, then there is a two-photon (2γ) formulation for the $dE/dx$ of magneticons. And for this scattering interaction, one has (the possibility of) two topologically distinct Feynman diagrams:[39] Fig. 1. Using the results of Appendix A, it is estimated that the $dE/dx$ for magneticons in the 2γ formulation is roughly twice that found in prior (1γ) formulations.[40] This augmentation of the ionization loss is due to the (provisionally assumed) presence of a second, or magnetic, photon. This result, if it is indeed the case, would be another feature of an experimental magneticon signature.

In view of this new result (possible additional ionization loss by magneticons), we examine the experiments that use ionization loss as a particle identifier from both points of view, i. e., with and without a factor of two. In existing $e^+e^-$ collider data, only in the Free Quark Search [28] does the conclusion depend upon the assumption of a 1γ or a 2γ formulation. (Note that this additional ionization does not figure into the Lorentz force tracking calculations for magneticons moving in a magnetic field.)

B. Existing $e^+e^-$ collider data

Eq. (22) indicates that for beam energies sufficiently above the limit posed by the magneticon pair mass, magneticon pairs should have been copiously produced at prior $e^+e^-$ collider experiments. Hence, it becomes of interest to investigate what experimental limits might, in the context of this paper, presently exist for magneticon pair production. First, we note that all of the major detectors at PEP,

PETRA, Tristan, SLC, and LEP were configured with large solenoidal magnetic fields to enable momentum analysis of the (electrically) charged particles in the final state reaction products. In these detectors, magneticons would be accelerated, or decelerated, by the detector's magnetic field rather than bent by it. This means that the tracks of these putative magneticons in the projection into a radial view, in which the *z*-axis is a central point, would be straight radial lines (rather than segments of circles), and that the track projections into the plane containing the *z*-axis and the magneticon track would be parabolas (in non-relativistic approximation). In this way, the standard all-purpose $4\pi$ detector design militates against the detection of magneticon pairs unless the trigger and tracking algorithms are specifically designed with such a possibility in mind.[41]

1. Magnetic Detectors

We are aware of three $e^+e^-$ collider experiments that actually did pursue the detection of the tracks of magnetic particles in the presence of magnetic fields. Examining these efforts, going from lower to higher CM energies, we note that the CLEO experiment at CESR [29] (with a 1 T magnetic field) reported an exclusion of monopoles with masses up to 5 GEV/$c^2$ and with charge $g \geq 2e$. Since their detection efficiency drops off rapidly for low magnetic charge, they evidently would not have seen magneticons with a charge of $g = e$. In any case, it is argued below that this range of magneticon mass (of charge $g = e$) is excluded by an experiment at DORIS II [30], which did not have a magnetic field.

We next consider the TASSO experiment at PETRA [31], which employed a 0.5 T solenoidal field. They looked for tracks of monopoles with a range of magnetic charge, reporting limits on $g = 137e/2, 60e, 50e, 40e, 30e, 20e$, and $10e$ in a mass range of $1 \leq m_\mathrm{m} \leq 16.5$ GeV/$c^2$, where this upper mass limit applies to $g \leq 30e$. (The effective sensitivity to objects of higher magnetic charge extends to magnetic objects of lower mass.) To understand the possibilities for detection at TASSO for $g < 10e$, examination of their detection efficiency plot indicates that they have 0% detection efficiency for $g = 5e$ for masses up to ~12 GeV/$c^2$, and 0% for $g = e$ covering all masses up to their highest sensitivity of ~ 16.5 GeV/$c^2$. Thus, the conclusion here is that TASSO, using their described trigger and analysis, would not have detected magneticons having a charge of $g = e$, even with an augmentation in the ionization loss, which would be associated with the $2\gamma$ formulation for the *dE/dx* of magneticons.[42]

Finally, the third of these experiments, the OPAL detector at LEP2 [32], looked for "classical" magnetic monopoles in a solenoidal magnetic field of 0.435 T. By classical is meant an object of large magnetic charge, though their analysis still included magnetic charges well below the Dirac charge of $137e/2$. While it is difficult to determine precisely how low a magnetic charge would pass their monopole cuts, we note that their monopole trigger was based upon a large amount of ionization energy loss as recorded in their jet chamber. Specifically, they required an integrated signal above a threshold of 1250 counts in the Flash Analogues to Digital Converters (FADC). Noting that a minimum ionizing particle would be expected to yield ~200 FADC counts (implying that twice minimum ionizing would yield ~400 FADC counts – still too small for their trigger criterion), it is fair to conclude that this OPAL monopole trigger requirement will preclude the detection of any magneticons of unit magnetic charge, which would be expected to have ionization losses on the order of minimum ionizing, or, as we argue below, ~twice minimum ionizing. Hence the limits reported by Ref. [32] do not bear upon the mass exclusion range for magneticons of charge $g = e$, as contemplated in this paper.

2. Non-magnetic detectors

There remains another fruitful avenue to explore: there were several smaller specialized detectors at $e^+e^-$ colliders that did not use magnetic fields for momentum analysis. The data from the non-magnetic detectors at lower energies enable the exclusion of magneticon masses from zero up to some upper limit set by the beam energy (less certain restrictions and corrections). A useful data set for this purpose is that taken by the Crystal Ball experiment at a beam energy up to ~5.29 GeV [half of the mass of the $\Upsilon(4S)$] at DORIS II at DESY [30].[43] In particular their data on the μ pair production cross section in the region of the $\Upsilon$ resonances, which is comprised of the direct $e^+e^- \to \mu^+\mu^-$, the $e^+e^- \to \Upsilon \to \mu^+\mu^-$, as well as the interference term. Here we consider the highest beam energy point (5.29 GeV) in their Fig. 1, which has the most leverage on the magneticon mass limit, and make the assumption that a magneticon contribution to the observed μ pair production rate in the amount of $\beta^3 \leq 2\%$ of a unit of $R$ would not be noticed. Again using Eq. (22), we find that $\beta = 0.27$ gives this 2% contribution to $R$, setting an upper limit to the magneticon exclusion zone for this experiment at $m_m <$ ~5.1 GeV/$c^2$.[44]

Looking at higher collider energies, there are three non-magnetic experiments at PEP: Anomalous Single Photon Search (ASP) [35], Search for Highly Ionizing Particles (monopole search) [36], and Free Quark Search (FQS) [28].[45] ASP looked for a gamma at $\theta_\gamma > 20°$ with no ionizing tracks in the rest of the detector. That is, the ASP trigger precluded the detection of magneticon pair production. The monopole search, using the plastics Lexan and CR-39 (also called nuclear track detectors, or NTDs), reported monopole charge limits in the range $20e \leq g \leq 200e$, precluding their sensitivity to magneticons carrying the above described magnetic charge $e_m$. Similar to Ref. [36], a number of monopole searches using NTDs were also undertaken at other accelerators, including hadron machines. (See the Particle Data Group Reviews [38].) Again, as in Ref. [36], they also would not have been sensitive to magneticons of unit charge. It is also appropriate to note here that NTDs have been used extensively to look for highly charged (Dirac) magnetic monopoles in cosmic rays.[46] None were found; these references can also be found in Ref. [38].

Of the three non-magnetic PEP experiments, then, only the exclusive FQS (beam energy = 14.5 GeV), which looked for pairs of back-to-back ionizing tracks consistent with a charge of $e/3$ and/or $2e/3$, could yield a discernable magneticon signature, but only within a certain range of magneticon mass. The excluded mass range in this re-analysis of their data depends upon the formulas used for magneticon $dE/dx$, which we have developed in Appendix B. There we first discuss formulations of $dE/dx$ for electrically charged particles, which would be what we have called the 1γ formulation. The 1γ formulation includes all of the $dE/dx$ (or $dE^{1\gamma}/dx$) analyses up to now because they tacitly accept as a given that there is extant only one type of photon – the electric photon. The Feynman diagram relevant for (the lowest order) one-photon e-P scattering is given by Fig. 3. This is the scattering diagram through which charged particles experience (most of) their energy loss as they pass through matter.[47]

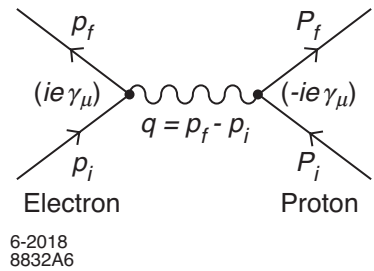

Fig. 3. Feynman diagram for e-P scattering. Only one photon, the electric photon is relevant. The QED vertex factors are given in parentheses. As discussed in the text, the labels MS and LF are not relevant.

An excellent physical discussion of charged particle collisions, energy loss, and scattering has been given by Jackson [15, Ch. 13]. And Fano [39] has published an excellent *dE/dx* review article. In addition, a comprehensive review of the history of *dE/dx* derivations has been given by Ahlen [22], who considered the details of the energy loss question for magnetic as well as electric particles. The analysis in Appendix B uses these references as well as the results of Appendix A to develop both a 1$\gamma$ and a 2$\gamma$ formulation for the *dE/dx* for magneticons. And it is pointed out in Appendix B that our results (for electrically) charged particles are in reasonable agreement with the the tables given by the NIST web site [73].

Using the formulae given in Appendix B, the magneticon energy loss results, as well as those for electrical particles (with charge $e$, $2e/3$, and $e/3$), are plotted (versus $\gamma - 1$) in Fig. 4, where it can be seen that the *dE/dx* lines for magneticons (for both the 1γ and 2γ formulations) intersect those for fractionally (⅓ and ⅔) charged quarks. [Using $(\gamma - 1)$ as the abscissa in Fig. 4 is possible because the *dE/dx* formulae do not depend explicitly upon the mass (only upon the velocity) of the projectile, thus enabling the elimination from the plot the functional dependence upon projectile mass.[48]] At the indicated intersection points, the putative magneticons and quarks will have the same $\gamma$ (and $\beta$) factors *and* the same (estimated) ionization rates. Thus, Fig. 4 demonstrates that the experimental exclusion by Ref. [28] of the existence of fractionally charged quarks below a mass limit of ~14 GeV/$c^2$ can be re-interpreted as also excluding magneticons, but with a different mass exclusion range. The reason for the different mass exclusion range(s) for magneticons is because of the differences in *dE/dx* as a function of $(\gamma - 1)$ for (electrically charged) quarks and (magnetically charged) magneticons – the 1γ and 2γ formulation each leading to its own mass exclusion range, as will be seen below.

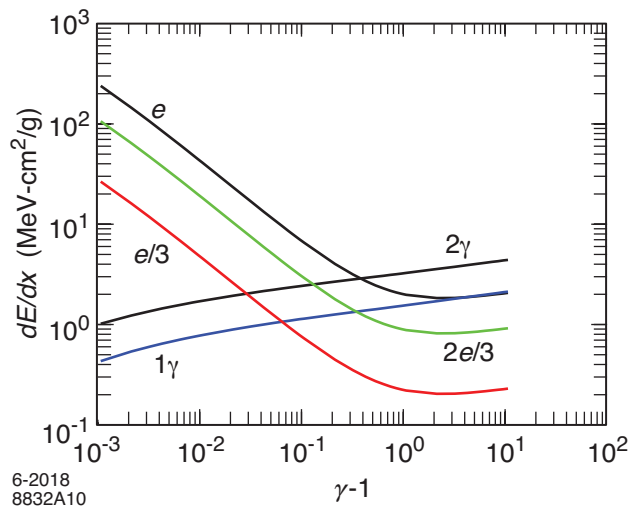

Fig. 4. *dE/dx* plots versus ($\gamma - 1$) for electric particles of charge $e$, $2e/3$, and $e/3$ penetrating carbon (to represent the stopping power of scintillator). Also given are the *dE/dx* plots, based upon the 1$\gamma$ and 2$\gamma$ formulations for magneticons penetrating carbon. Note that since the *dE/dx* formulas don't explicitly contain the projectile mass, the *dE/dx* curves in Fig. 4 are universal curves (except for the electron, and other possible particles of comparable , or less, mass).

To continue our re-interpretation of the FQS data, we examine Fig. 2 of the FQS paper [28] [reproduced here as Fig. 5 (with some additional $Q$ lines, as explained below)]. Fig. 5 presents a scatter plot of ~13,000 undifferentiated back-to-back events (mostly Bhabhas): on the abscissa is the projected distribution of deduced charge in Arm 1 (referred to herein as $Q_1$), and on the ordinate that quantity in

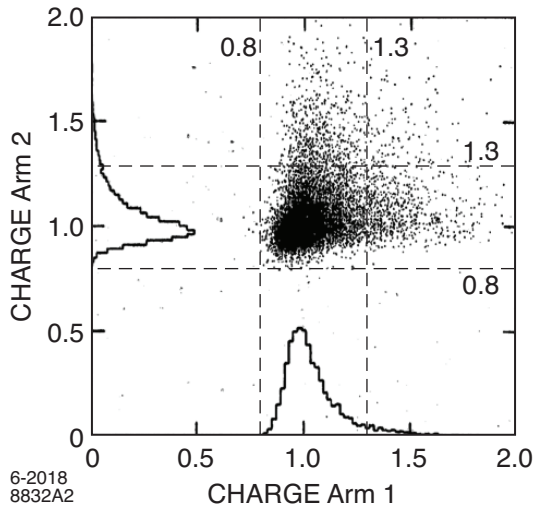

Fig. 5. FQS scatter plot (their Fig. 2) of the calculated charges of ~13,000 back-to-back pair events. This plot also includes projections of the events onto the two axes of the plot. We have drawn in lines at $Q = 0.8$ and $1.3$, which, as further described in the test, suggest boundaries for the use of this data for the purpose of detecting magneticon pairs.

Arm 2 ($Q_2$). (The $Q$ scales of Fig. 5 are the same as the $Q$ values of Fig. 4.) Each pair produced final state will ideally appear as a point on the $Q_1 = Q_2$ diagonal line. As expected, this scatter plot distribution has a maximum (due to pair produced electric particles) centered at $Q_1 = Q_2 \sim 1$ – but with a spread due to experimental measurement error in ToF[49] and ionization (the latter being further exacerbated on the high side by the existence of Landau tails). It is clear that in Fig. 5 the pair produced magneticon final states would also be expected to appear as points on (or near) the diagonal $Q_1 = Q_2$ line, being scattered from their ideal locations by experimental error in the *dE/dx* and ToF measurements. The magneticon signature (reduced *dE*/*dx*), then, derives from the fact that slow magneticons would mimic the anticipated ionization loss of electrical particles with reduced electric charge, which is exactly the quark (or fractionally charged lepton)[50] signature in the FQS.

At this point we explore in more detail the FQS derivation of the deduced charge $Q$ from the measured energy loss (rate) in the experimental apparatus, which we denote here by $dE_{exp}/dx$. For the FQS, the experimentally measured stopping power is compared to the usual stopping power formula for charged particles, e. g., our Eq. (B18), which we rewrite here in a suitable functional form as

$$\frac{dE_Q}{dx} = \frac{Q^2}{\beta^2} F(\beta): \qquad (24)$$

as in the FQS[51], we have introduced the (fractional) quark charge $Q$, and have extracted the $\beta^2$ factor (in the denominator) from Eq. (B18), including the rest of the function as $F(\beta)$. Thus, we obtain:

$$Q = \beta \sqrt{\frac{dE_{exp}/dx}{F(\beta)}} \qquad (25)$$

as (our version of) the FQS charge assignment algorithm.[52] In our effort to re-interpret the FQS data to be relevant to magneticons, we rewrite Eq. (25) as:

$$Q = \beta \sqrt{\frac{dE_{em}^{n\gamma}/dx}{F(\beta)}}, \qquad (26)$$

where we have replaced the $dE_{exp}/dx$ by the magneticon energy losses, as derived in Appendix B. [We set $n\gamma = 1\gamma$, Eq. (B21), or $2\gamma$, Eq. (B22), as appropriate for the two formulations for magneton energy loss.] The $\beta$ and $F(\beta)$ remain in Eq. (26) because they are still a part of the charge assignment algorithm for the individual events. As an adjunct to Fig. 4, using Eq. (26), we plot Fig. 6, which indicates the relationship (in both formulations) between magneticon mass (which, given the fixed beam energy of 14.5 GeV, uniquely determines its $\beta$) and its charge $Q$, as assigned by our algorithm. As suggested boundaries to a detectable magneticon pair signature (as explored below), in Fig. 6 we have drawn lines at $Q = 0.8$ (1.3) for low (high) $\beta$, high (low) mass magneticon pairs.

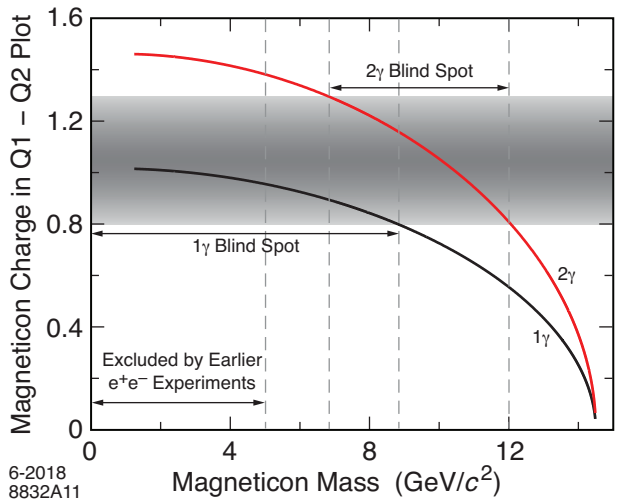

Fig. 6. Plot indicating the value of $Q$ that would be assigned, as a function of magneticon mass, to a magneticon by the FQS back-to-back quark pair analysis. A line is plotted for each *dE*/*dx* formulation. Also, as in Fig. 5, are plotted lines at $Q = 0.8$ and 1.3, which indicate approximate boundaries to the use of the FQS data to detect magneticon pairs, and which are indicated in Fig. 6 for the two formulations. The shading between these lines indicates that there are blind spots for the use of the FQS data for the detection of magneticon pairs. Also, a vertical line at 5 GeV/$c^2$ is drawn in as an indication of the upper liimit of magneticon mass already deemed to be excluded by early non-magnetic detectors at $e^+e^-$ colliders.

There are two approaches to discerning a magneticon pair signature in the FQS data: (1) look for a bump or a shoulder in the projection histograms along the $Q$ axes of Fig. 5, or (2) examine the scatter plot itself for an unexpected accumulation of pair events (along the diagonal line).

Because of the small number of expected magneticon pair events, we see that the first approach is disadvantaged should those pair events be anywhere within the skirt regions of the projections of the scatter plot events. As an example, at a magneticon mass of 8.8 GeV/$c^2$ (which, for the 1γ formulation, Fig. 6 indicates a $Q = 0.8$) and taking the appropriate $\beta^3$ factor into account, one anticipates on the order of a half of the expected μ pair total of ~460 events (see Appendix C). Thus, assuming an equal experimental width for the projected magneticon distribution, we argue that a 230 magneticon pair distribution, when projected onto either $Q$ axis, will have a peak of height 230/13,000 times that of the main distribution, that is, ~0.01 on the $Q$ scale. (For this calculation, we note that in Fig. 5 the peaks of the projections of the main distribution, which contains ~13,000 events, measure out at ~0.5 on the $Q$ scale.) This height is roughly comparable to the width of the lines used to draw these graphs. Hence, unless such a magneticon distribution would be beyond the skirts of the main distribution, it would be submerged in the main distribution of electrical particles of unit charge, and would probably not be noticed. Since for $Q < 0.8$ the scatter plot in Fig. 5 is essentially clean, we conclude that either approach would be possible. Thus, as indicated in Fig. 6, we have indicated a blind spot for the 1γ formulation to extend from 0 to ~8.8 GeV/$c^2$. Using the same argument, but taking into account the larger ionization of the 2γ formulation, we have set in Fig. 6 the upper boundary of the blind spot of the 2γ formulation (also associated with $Q = 0.8$) at ~12 GeV/$c^2$.

Looking toward higher masses (i. e., lower $\beta$s), the upper magneticon mass exclusion boundary for the FQS is firmly set by the beam energy of 14.5 GeV. However, as a practical matter, a mass of 14.5 GeV/$c^2$ would have $\beta = 0$ and thus have a null pair production probability. To take this fact and the reduced magneticon ionization into account, we provisionally adopt the mass limit of 14 GeV/$c^2$, which is the same as that given by the FQS for fractionally charged leptons. We note, however, that because of the lower magneticon pair production cross section (which goes like $\beta^3$), this limit is not as robust as the FQS limit for fractionally charged leptons or quarks. For a 14 GeV/$c^2$ magneticon mass, the estimated number of event pairs expected in the solid angle of the FQS detector (not counting trigger and tracking losses) is ~8 events.

We now observe that for the high $Q$ region of Fig. 5, the Landau tails extend the projections of ~13,000 events even beyond $Q_1$ and $Q_2 \sim 1.5$. (This is even beyond the highest magneticon charge assignment by the 2γ formulation, a nominal $Q \sim 1.4$.) Hence, in this case, to try to improve the range of magneticon pair detectability, one turns to approach (2) and looks for an accumulation of events on the diagonal line of the scatter plot. It appears to us that in Fig. 5, a distribution of ~400 events ($\beta$ is somewhat larger for these lighter particles) would probably be noticed on the scatter plot centered at $Q_1 = Q_2 = 1.5$, but would be moving into the main distribution near a value of $Q \sim 1.3$. But since we are seeking nominal limits here, we refrain from arguing that such a distribution might be noticed at a smaller $Q$, and set the lower limit of the 2γ formulation at the associated magneticon mass of ~6.7 GeV/$c^2$. Recognizing that the other non-magnetic experiments have precluded magneticons of mass < 5 GeV/$c^2$, we find that the FQS data extends this exclusion range up to ~14 GeV/$c^2$, but with with significant blind spots.

Before leaving this discussion of the FQS data, we observe that in Appendix C, we calculated that, for the reported integrated luminosity (15.5 ±0.4) pb$^{-1}$ and solid angle 4π/3, one expects ~460 μ pair and ~7600 Bhabha events in the $β \sim 1$ peak of the scatter plot. It is interesting to note that the sum of these estimated contributions, ~8060 pairs, is significantly less than the ~13,000 pair events recorded. Of course, there are a number of possible corrections and additional backgrounds that might be applicable to this data set, but these are difficult to evaluate without more detailed information about the experiment. (Off hand, though, there certainly seems to be room for ~400 magneticon pair events to be submerged in this peak; in a follow-on experiment, it would be useful to have good particle ID to assist in sorting this question out.)[53] It should also be said, however, that the apparent surplus of events in this peak does not vitiate the original limits placed by the FQS on the existence of fractionally charged particles, which derive from the empty region below the unit charge peak.

Finally, in concluding this discussion of e$^+$e$^-$ collider data, it seems safe to say that below $m_m$ = about 5 GeV/$c^2$, one or more of the prior experiments would probably have seen a magneticon signal. However, in view of the hazards in re-analyzing prior experiments (in particular the FQS) for an unanticipated and unsought signal using ionization formulae that are not experimentally verified, it would seem appropriate to recommend a specific search for a magneticon signal for $m_m > 5$ GeV/$c^2$.

C. Induction technique and superconducting quantum interference device (SQUID) detectors

Brief mention should be made of the use of the induction technique for the detection of magnetic monopoles. This technique was pioneered by Alvarez, et al. [58], who circulated moon rocks many times through a superconducting coil to build up a cumulative (current) signal for magnetic charge. They subsequently refined their apparatus, using SQUID technology (achieving a much lower background and a much greater sensitivity) to continue their searches, also looking for monopoles, possibly stopped in various accelerator components, e. g., beam pipes [62]. The SQUID technique was continued by a number of other experimental collaborations, looking for (Dirac) monopoles in a variety of substances, e. g., Kovalik and Kirschvink [63]. Recent SQUID experiments have been published by, e. g., Kalbfleisch, et al. [59], and by Atkas et al. (H1 Collaboration) [60] and by Acharya et al. (MoEDAL Collaboration) [67], who looked for possible stopped monopoles in various components of accelerator equipment. Ref. [59] reports an rms spread in data run step size of 0.73/2.40 (~0.3 of a Dirac monopole), while Ref. [60] claims a sensitivity down to 0.1 of a Dirac monopole. In their samples Ref. [67] excluded monopoles of charge $|g| \geq 0.5 g_D$, but was insensitive to smaller $|g|$. To date no (Dirac) monopoles have been found [38].

Ref. [64] indicates that with typical parameters, a SQUID detector can achieve an intrinsic rms noise level of ~$1.5 \times 10^{-7} \Phi_0/\sqrt{Hz}$, where $\Phi_0 = hc/2e$ is the unit quantum of magnetic flux, twice the flux that would emanate from a Dirac monopole. (At this sensitivity a SQUID detector could easily discern a single magneticon.) However, practical devices, especially those with warm bores to accommodate physical samples, tend to sustain much higher noise levels. It is tantalizing to observe that if there would be accelerator produced magneticons residing in the samples of Refs. [59 and 60], their experimental rms spreads could easily include a sizeable number (a dozen or more) of stopped magneticons.[54] It is clear that serious attention to experimental background noise and systematics would be required to obtain a reliable SQUID signal from singly charged magneticons.

Another series of SQUID searches for magnetic monopoles was initiated by the spectacular monopole candidate signal found by Cabrera [61]. This candidate was presumably a heavy, slow moving Dirac monopole from a cosmic source.[55] The numerous subsequent cosmic ray searches, using SQUIDs and other techniques, have increased the integrated time-area sensitivity (over that obtained by the original Cabrera experiment) by many orders of magnitude. To date, no monopole has been found [38], leading the conclusion that the original monopole candidate signal of Cabrera was some pernicious background event (albeit with exactly the right magnitude for a Dirac monopole).

D. Cosmic ray showers[56]

In looking at singly charged magneticons of mass above 5 GeV/$c^2$, one might look to cosmic ray experiments; there is certainly plenty of available energy. However, one must recall that the pair production cross section by a gamma ray impinging on a nucleus goes like $m^{-2}$ [45]. Thus, looking at a possible lowest mass magneticon as determined by our above analyses, one estimates that the cosmic ray gamma production cross section of 5.1 GeV/$c^2$ magneticon pairs will be a factor of ~$10^8$ below that for electron pairs. And the production cross section of heavier magneticon pairs will have an even larger suppression factor. Following the above line of argument indicates that the production of possible magneticons of mass > 5 GeV/$c^2$ would have most certainly escaped notice in cosmic ray experiments.

E. The LHC

It is possible to look in the LHC data (as well as that from other hadron machines) for magneticon pairs. Based upon the analysis above, one can easily estimate an expected production rate from at least one hadronic interaction – the Drell-Yan process [46]: to wit, a quark-antiquark annihilation produces a virtual photon that in turn decays to a lepton-antilepton pair. The usual Drell-Yan process is analogous to that depicted in Fig. 2a, but with quarks annihilating at the initial vertex. And the anticipated Drell-Yan magneticon pair production would be analogous to that depicted in Fig. 2b, also with quarks annihilating at the initial vertex. As indicated in the text, then, sufficiently above threshold, the m$\bar{\text{m}}$ production rate would equal that for the Drell-Yan $\mu^+\mu^-$ production, which is quite copious up to and well beyond $m_{\mu\bar{\mu}}$ = 260 GeV/$c^2$ [47, 48]. Except for the magnetic bending, the m$\bar{\text{m}}$ events would resemble the $\mu^+\mu^-$ events. Of course, to obtain these m$\bar{\text{m}}$ events, one would have to fashion suitable magneticon trigger and tracking algorithms. Otherwise, even if present in copious quantities, these $m\bar{m}$ events would not be observed. We argue that further investigation of this possibility is warranted [68, 69].[57]

F. Future colliders

Similarly, for possible future observation of magneticons, if suitable tracking and trigger algorithms were written, one could look to future $e^+e^-$ machines, i. e., a Super B Factory [49],[58] an ILC [50], CLIC [51], or Higgs factory [52], which would afford at high energies a far cleaner situation than that at the LHC or other hadron machines. The higher energy beams might also produce second generation magnetic leptons and/or also heavier, presumably "hadronic," magneticons.

G. Anomalous magnetic moments of the electron and muon

An important question to ask at this juncture is: Would the existence of the lightest permissible charged magneticon, as determined by the above analysis (which for the purpose of the discussion of this section we take to have $m_m = 5.1$ GeV/$c^2$), have an effect large enough to be relevant to the best theoretical calculations and/or existing measurements of the anomalous magnetic moment $a_e$ of the electron or $a_\mu$ of the muon? Also, what would be the effect of the $2\gamma$ formulation? To set up a framework in which to answer these questions, for the lepton $\ell$, we first make the following decomposition for the theoretical calculations (using the SM):

$$a_\ell^{SM} = a_\ell^{QED} + a_\ell^{wk} + a_\ell^{had}, \qquad (27)$$

where, as in Ref. [53], $a_\ell^{QED}$, $a_\ell^{wk}$, and $a_\ell^{had}$ are the QED, (electro) weak, and hadronic contributions, respectively.

Based upon the concept of the existence of a magnetic world, as presented in this paper, (at least) two additional terms, $a_\ell^{QEMD}$ and $a_\ell^{L \times L}(\ell, m)$ representing magneticon contributions to $a_\ell$, might be appended to Eq. (27). These contributions arise from the presence of closed magneticon loops, analogous to the closed fermion loops contributing to $a_\ell^{QED}$. For example, Fig. 7 gives the lowest order (i. e., a two loop diagram) QED Feynman diagram (in momentum space) for $a_\ell^{QED}$ that contains a vacuum polarization (VP) loop of a heavier fermion, e. g., a tau. Fig. 8 is the diagram that would be associated with the analogous QEMD magneticon VP loop; in this figure, using the arguments presented above, we indicate in parentheses the appropriate vertex factors. We view the photon path in Fig. 8 as an *s*-channel (or time-like) path, analogous to the $m\bar{m}$ pair production discussed above. Fig. 9 schematically shows three (of six) QED light-by-light diagrams. In this case, the closed fermion loop would be a tau. Fig. 10 shows in more detail one of the (tau-loop) diagrams depicted in Fig 9. Fig. 11 is like Fig. 10, except that it represents a closed light-by-light magneticon loop. We view the essence of this diagram as a *t*-channel (or time-like) path, in that it finds a conceptual analogue in Fig. 1b. There is also the two photon question, the effects of which we shall discuss in additional detail in Appendix A.

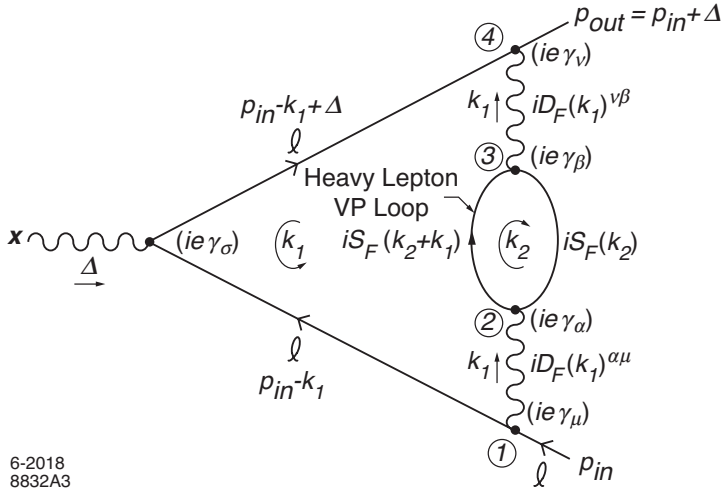

Fig. 7. Feynman diagram in momentum space for the lowest possible order ($n = 2$) of a vacuum polarization contribution (by a heavy lepton VP loop, e. g., a tau for conventional QED) to a light lepton (e or µ) electromagnetic vertex (indicated by **X**). For the photon and its VP loop, the vertices are numbered 1 through 4 in sequence. We have indicated (in parentheses) the QED interaction vertex factors, $ie\gamma_\mu$, for each vertex. (Recall that they are positive because we take $e > 0$, as discussed in the text.)

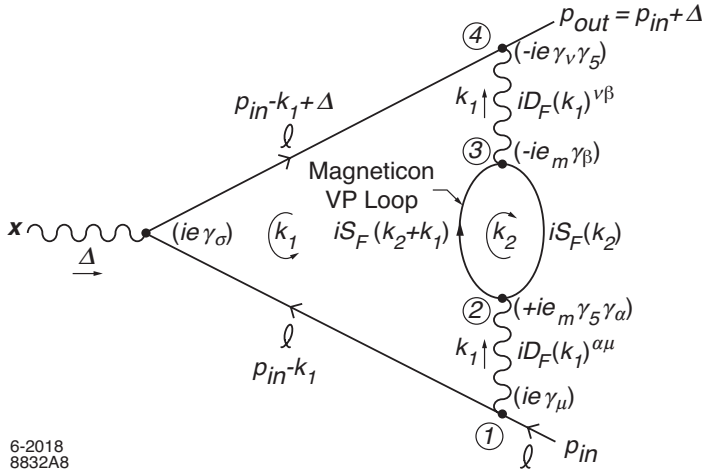

Fig. 8. The Feynman diagram for a closed magneticon VP loop analogous to the heavy lepton VP loop in Fig. 7. For the analogous QEMD diagram, (based upon the discussion in the text) we also have indicated in parentheses at each vertex the appropriate QEMD vertex factors when the VP loop consists of a magneticon.

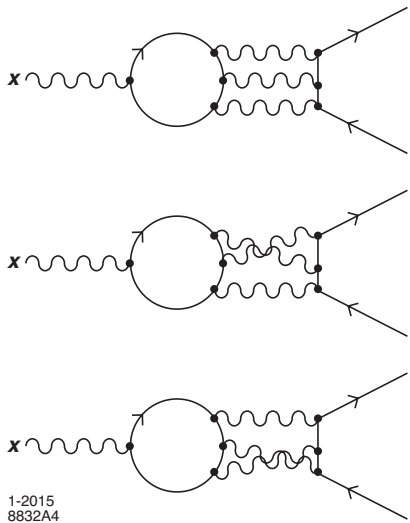

Fig. 9. Three (of the 6) Feynman diagrams (associated with a closed fermion loop) for the light-by-light contribution to the anomalous magnetic moment of a lepton. The other three diagrams are obtained by reversing the direction of circulation of the fermion in the loop, which is indicated by an arrow on the loop.

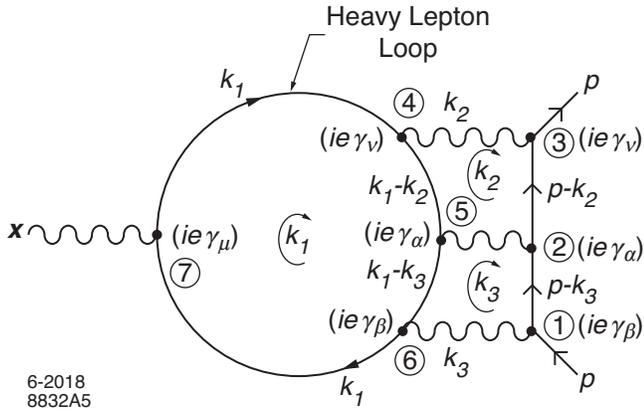

Fig. 10. One of the six Feynman diagrams in momentum space for the QED light-by-light Feynman diagrams which contains a closed heavy lepton (e. g., a tau) loop. The 3 vertices along the light lepton track, and the 4 vertices in the tau loop are numbered in sequence. As in Fig. 7, the vertex factors are indicated in parentheses.

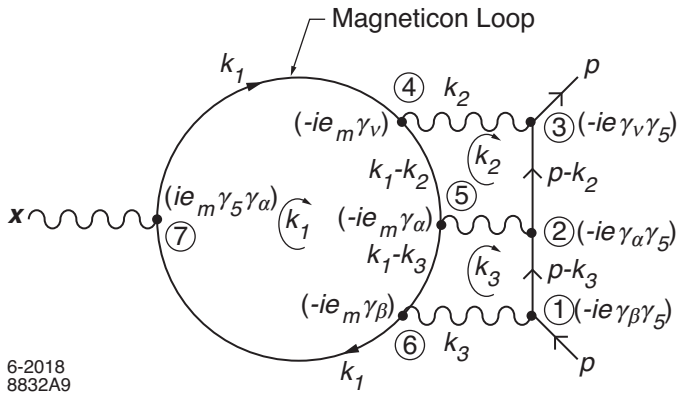

Fig. 11. One possible QEMD light-by-light Feynman diagram analogous to Fig. 10, but with a closed magneticon loop. One possible set of QEMD vertex factors (all magnetic photons), which are consistent with the $t$-channel scattering diagram as depicted in Fig. 1b, are indicated in parentheses. As discussed in the text (and as can be seen in Fig. 1a), each of these magnetic photons can also replaced by electric photons, multiplying the number of diagrams by a factor of $2^4 - 1 = 15$. As indicated in Fig 1, the vertex factors will be changed accordingly.

Having the vertex factors, propagators, charge magnitudes, etc., one could write down and evaluate in the usual way the formulae associated with these magneticon diagrams. However, in Appendix A we take a simpler but straightforward approach: we argue that the mathematical expressions that represent these QEMD diagrams (Figs. 8 and 11) can be related to those associated with the analogous QED diagrams in which the fermion loop would contain a heavy lepton, formulae for which have already been derived. Thus, using this relationship, a QEMD evaluation can be done using the mass of the proposed magneticon (in place of that of the tau) in the appropriate QED formulae. (For this purpose, Knecht [54] furnishes a useful compilation of relevant formulae.) We show that the final mathematical expression for $a_\ell^{\text{QEMD}}$ leads to the same numerical result as $a_\ell^{\text{QED}}$, and conjecture that the same arguments also hold for the $a_\ell^{\text{L}\times\text{L}}(\ell,m)$ diagrams. (For the 2γ formulation there is also a topological factor; see endnotes [59] and [60].)

Looking first at $a_e^{\text{SM}}$, as the physics analysis of this quantity presently stands, it is known that the dependence of $\alpha$ on any contribution other than $a_e^{\text{QED}}$ in Eq. (27) is negligible [53]. And, based upon the logic and calculations given in Appendix A, we argue that the proposed new magneticon terms, $a_e^{\text{QEMD}}$ and $a_e^{\text{L}\times\text{L}}(e;m)$ do not change this conclusion. That is, for the electron, using the values for $m_e$, as given by the PDG and $m_m$ = 5.1 GeV/$c^2$, the lightest possible magneticon, we obtain $a_e^{QEMD} = 1.2 \times 10^{-15}$. Similarly, in the 1γ formulation, we obtain for the lowest order magneticon light-by-light contribution $a_e^{L \times L}(e;m)_{1\gamma} = 7.7 \times 10^{-17}$. One can see that for the 2γ formulation, even with a topological augmentation factor[59] of 15 for the magnitude of $a_e^{L \times L}(e;m)_{2\gamma}$ is still projected to be too small to alter this conclusion.

In Table I we collect the values of $a_e^{SM}$ and its components, the best experimental measurement for $a_e^{exp}$ [55], as well as $a_e^{QEMD}$ and $a_e^{L \times L}(e;m)$. (N. B., for consistency, we have used the value given in Ref. [53] for $\alpha$ and its uncertainty, i. e., $\alpha^{-1}$ = 137.035 999 074(44) to evaluate $a_e^{QED}$.)

**Table I.** Values for $a_e^{SM}$ and its components as compared to $a_e^{exp}$, to $a_e^{QEMD}$, and to $a_e^{L \times L}(e;m)$.

| Quantity | Value (uncertainty) | Uncertainty |
|---|---|---|
| $a_e^{QED}$ | 0.001 159 652 179 10(33) | $3.3 \times 10^{-13}$ |
| $a_e^{wk}$ | 0.000 000 000 000 029 73(52) | $5.2 \times 10^{-16}$ |
| $a_e^{had}$ | 0.000 000 000 001 685(22) | $2.2 \times 10^{-14}$ |
| $a_e^{SM}$ | 0.001 159 652 180 81(33) | $3.3 \times 10^{-13}$ |
| $a_e^{exp}$ | 0.001 159 652 180 73(28) | $2.8 \times 10^{-13}$ |
| $a_e^{QEMD}$ | 0.000 000 000 000 001 2 | $1.2 \times 10^{-15}$ (*) |
| $a_e^{L \times L}(e;m)_{1\gamma}$ | 0.000 000 000 000 000 077 | $7.7 \times 10^{-17}$ (*) |
| $a_e^{L \times L}(e;m)_{2\gamma}$ | 0.000 000 000 000 001 2 | $1.2 \times 10^{-15}$ (*) |
| (*) The evaluations for $a_e^{QEMD}$ and $a_e^{L \times L}(e;m)$ are included in the last column for easy comparison. These particular numbers are magnitudes – not uncertainties. The subscripts 1γ and 2γ indicate the number of fundamental photons used in the evaluations. | | |

As can be seen in Table I, while both $a_e^{wk}$ and $a_e^{had}$ contribute (small amounts) to the final value of $a_e^{SM}$, the errors in these values are not significant relative to the uncertainty in $a_e^{QED}$. Also, it is evident that the magnitudes of the estimates for $a_e^{QEMD}$ and $a_e^{L \times L}(e,m)_{1\gamma}$ and $a_e^{L \times L}(e,m)_{2\gamma}$ are much too small[60] to be relevant to the evaluation of $a_e^{SM}$. Thus, the agreement between $a_e^{SM}$ and $a_e^{exp}$ (because the value of $\alpha$ is used to calculate $a_e^{QED}$ which in turn derives from the value of $a_e^{exp}$) remains in force.

We now turn to $a_\mu^{SM}$. In contrast to the agreement between $a_e^{SM}$ and $a_e^{exp}$, in the case of $a_\mu$ there is, at present, a discrepancy between theory and experiment in the range of 3.2 to 3.6$\sigma$, depending upon the theoretical estimates used for the hadronic vacuum polarization contribution [56]. The difference between the electron and muon calculations is due to the fact that the muon is ~200 times heavier than the electron, which gives the heavy fermion (tau) loop corrections a much larger role in the evaluation of $a_\mu^{SM}$. The logic developed in Appendix A for $a_e^{QEMD}$ can also be used to calculate $a_\mu^{QEMD}$. Using the PDG value for $m_\mu$ and $m_m$ = 5.1 GeV/$c^2$ we obtain $a_\mu^{QEMD}$ = 5.14.×10$^{-11}$. Similarly, following the logic given in Appendix A, we obtain $a_\mu^{L\times L}(\mu;m)_{1\gamma}$ = 3.3×10$^{-12}$. Multiplying by a factor of 15 for the projected number of additional 2γ Feynman diagrams, we obtain $a_\mu^{L\times L}(\mu;m)_{2\gamma}$ = 5.0×10$^{-11}$. We collect in Table II these terms relevant to $a_\mu$.

**Table II.** Values for $a_\mu^{th}$ and its components (with uncertainties) as compared to $a_\mu^{exp}$

| Quantity | Value (uncertainty) | Uncertainty |
|---|---|---|
| $a_\mu^{QED}$ | 0.001 165 847 181 0 (15) | 1.5×10$^{-12}$ |
| $a_\mu^{wk}$ | 0.000 000 001 53 (1) | 1×10$^{-11}$ |
| $a_\mu^{had}$ | 0.000 000 069 29 (49) | 4.9×10$^{-10}$ |
| $a_\mu^{SM}$ | 0.001 165 918 01 (49) | 4.9×10$^{-10}$ |
| $a_\mu^{exp}$ | 0.001 165 920 89 (63) | 6.3×10$^{-10}$ |
| $a_\mu^{exp} - a_\mu^{SM}$ | 0.000 000 002 88 (80)   (3.6$\sigma$) | 8.0×10$^{-10}$ |
| $a_\mu^{QEMD}$ | 0.000 000 000 051 4 | 5.14×10$^{-11}$  (*See Table I) |
| $a_\mu^{L\times L}(\mu;m)_{1\gamma}$ | 0.000 000 000 003 3 | 3.3×10$^{-12}$  (*See Table I) |
| $a_\mu^{L\times L}(\mu;m)_{2\gamma}$ | 0.000 000 000 053 | 5.0×10$^{-11}$  (*See Table I) |

While the estimated value of $a_\mu^{L\times L}(\mu;m)$ is too small to be relevant for the evaluation of $a_\mu^{th}$, it can be seen that if $m_m$ is 5.1 GeV/$c^2$, $a_\mu^{QEMD}$ will induce a small closing of the gap between $a_\mu^{th}$ and $a_\mu^{exp}$. Specifically, in the 1γ formulation, the gap between theory and experiment would close to from (288 to 283)×10$^{-9}$, equivalent to a small improvement – from 3.6$\sigma$ to 3.5$\sigma$. And in the 2γ formulation, the gap between theory and experiment would close from (288 to 278)×10⁻⁹, reducing the discrepancy to 3.4$\sigma$. And if indeed there were additional particles in a so-far unobserved magnetic sector, one could expect additional terms to be relevant to this calculation, further closing the gap. Is it conceivable that the sum of the contributions of a magnetic sector might fully close this gap?

## V. SUMMARY AND CONCLUSIONS

The foundation for the physics of this paper is a generalization of the symmetries of classical electromagnetism to include magnetic charges and currents [6]. This generalization is formulated in such a way to exhibit a full dyality symmetry (see endnote [2]). Furthermore, we propose to extend this symmetry to quantum mechanics by incorporating it into a certain model for elementary fermions [3]. As a consequence, one predicts a set of magnetic counterparts to the set of SM (electric) fermions (both leptonic and hadronic). The lightest magnetic particle, which we call a magneticon, would be a (leptonic) counterpart to the electron[61]; it would be stable, have spin ½, have a magnetic charge equivalent to 1$e$, and have a Bohr electric moment, as dictated in the usual way by its charge and mass. Similarly, one would expect to have a full complement three generations[62] of spin ½ leptons and hadrons. (The existence of a set of hadronic bosons is also predicted to exist, but this is predicated on the assumption that the magnetic version of the hadronic interaction would enable the formation of suitable stable or metastable states.) Since no light magneticon has yet been seen, it is clear that dyality symmetry is broken. [Our view is that the broken dyality symmetry is in the mass attribute (magnetic particles are heavier than their electric counterparts), but not in the charge attribute, i. e., Eq. (23).][63]

We have examined extant experimental data and find that, to date, only non-magnetic detectors are relevant. Detectors using magnetic fields have not explicitly looked for magnetic charges as small as 1$e$ (In general, the Dirac charge, ~68.5 $e$, has been the central motivation for monopole searches.), and without explicitly looking for a 1$e$ monopole, such monopoles will not find their way into the data stream, generally speaking even for possible data mining off-line at a later date. (Magnetic bending is transverse to electric bending, and such monopole trajectories are discarded by the tracking algorithms.) Using data from non-magnetic detectors at e⁺e⁻ colliders, we find that magneticons of mass < 5 GeV/$c^2$ are excluded with a fairly high degree of probability. In addition, it appears that the FQS [28] restricts the existence of magneticons in some ranges of higher mass (but below the FQS beam energy of 14.5 GeV). However, the proper specification of these exclusion ranges is subject to interpretation, and our conclusion is that magneticon searches in the mass range from 5 GeV/$c^2$ up should be undertaken. It is our understanding that experimenters at SuperKEKB plan such a search for magneticon masses up to the energy limit of their collider [77]. To our knowledge, this is the first such explicit search for 1$e$ magnetons.

Classical electromagnetic interactions take place via electric and magnetic fields, that, in accordance with Maxwell's equations, are produced by electric charges and currents. In the analysis herein, we annotate this field production process as being associated with Maxwell source (MS) terms. (Electric particles are surrounded by electric and magnetic fields that the electric charges create.) In classical electrodynamics, these electromagnetic fields can be represented by the field tensor $F_{\mu\nu}$. In turn, these fields act upon other electric charges (and currents), as dictated by the Lorentz force equation via what we have called Lorentz force (or LF) terms. In quantum electrodynamics (or QED), the electromagnetic fields are represented by the photon which we view here as the "electric" photon, as it arises from quantizing the associated (electric) vector potential $A_\mu$. By the analogy of dyality symmetry, there also exists what we call magnetoelectricity, where the charges and currents are magnetic, and, by the magnetic version of Maxwell's equations, they would also generate electric and magnetic fields, but represented by a field tensor $G_{\mu\nu}$ (with its associated vector potential $M_\mu$, say). It was shown in Ref. [6] that $F_{\mu\nu}$ and $G_{\mu\nu}$ differ in parity.[64] Thus, in principal, they are distinct and physically distinguishable. From the magnetic (only) version of this theory, one could develop quantum magnetodynamics (or QMD), which would structurally be exactly like QED – in all regards except for the nature of the charge (and its associated fields), brought about by the application of a dyality rotation of $\Theta_d = \pm\pi/2$. In fact, it is noted that by re-defining the electron as magnetic, this is exactly the theory we would have (i. e., magnetoelectricity and its associated QMD), and we would have no way to tell the difference. These separate descriptions of quantum mechanics (QED *or* QMD) are consistent with present theory and lead to no contradictions because the underlying assumption is that there is only one kind of charge – either electric *or* magnetic (or at some fixed angle in the electromagnetic plane) – and only one kind of photon.

But neither (classical) electromagnetism nor magnetoelectricity, as described above, is dyality invariant: there is only one kind of charge. We now observe that Ref. [6], using a Clifford algebra formulation, joins these two theories together into a generalized electromagnetism that is dyality invariant.[65] There are both electric and magnetic charges (and currents), and these still generate electric and magnetic fields (i. e., $F_{\mu\nu}$ and $G_{\mu\nu}$) by the above described MS terms in conformance with electromagnetism, or magnetoelectricity, as described above. And these fields act on electric or magnetic charges via a generalized Lorentz force term, which contains the usual electromagnetic and magnetoelectric LF terms. Up to this point we have the usual physics expressions and expectations. But in our dyality invariant formulation for generalized electromagnetism, the generalized Lorentz force equation also has cross terms (which we label LFc terms), in which the electromagnetic fields ($F_{\mu\nu}$) act on the magnetic charges (and currents) and the magnetoelectric fields ($G_{\mu\nu}$) act on the electric charges (and currents). In this formulation, these cross terms contain a $\gamma_5$, which devolves from a $\gamma_5$ in the cross interaction terms of the Lagrangian. ($\gamma_5$, being a pseudoscalar quantity, enables the interaction cross terms of the Lagrangian to be expressed as scalar quantities, as is required for a proper Lagrangian.) In moving these generalizing concepts to a generalized quantum mechanics, which we label QEMD[66] (or QMED), we have proposed that we continue to hold dyality symmetry as fundamental. As a consequence, as shown in Ref. [5], there must be two photons, an electric photon *and* a magnetic photon.

In proposing to develop a viable QEMD (having two photons), we maintain that the analogy with the classical theory is important. Consequently, based upon a physical picture of charged particles, we argue that photon emission is always via MS terms and that photon absorption is always via LF terms. Now, as was pointed out by Bjorken and Drell [21], in QED this is an unimportant distinction; one gets the same answer either way one chooses to describe photon exchange between two (electrically) charged particles (and only one Feynman diagram is required). But at this point, we observe that in generalized electromagnetism, while the MS terms will remain the same, the LF terms will differ when electric charges interact with magnetic charges: that is, the emission vertex will always be the usual MS term, but the absorption vertex will be via an LFc term that will contains a $\gamma_5$. We propose to carry this distinction into QEMD and label the MS, LF, and LFc terms, as appropriate, and include a $\gamma_5$ in the LFc terms. (We note that the proviso "as appropriate"[67] needs experimental exploration and verification.) This argument leads in a natural way to the need for additional Feynman diagrams, as explored by calculation in the text. One basic and important consequence evolves from this formulation. It enables the postulation of the existence of a magnetic photon and, at the same time, gives a theoretical explanation for why we don't see any evidence for it: we live in an electric universe and, by the above argument using the appropriate MS terms, electrically charged particles can emit only electric photons. (While this picture appears to violate *T* invariance, it is clear that *TCP* or rather *TMCP* is still maintained (in the generalized electromagnetic equations).[68] Thus, we argue that the dyality invariant physics picture presented here satisfies the appropriate invariance relations.)[69]

Using our QEMD Feynman rules, we have made several calculations, which can be viewed as predictions of this theory. If correct, the observation and study of new physics beyond the SM is possible: for example, magneticon pair production by $e^+e^-$ collisons (or also by the Drell-Yan process in hadronic collisions),[70] and the energy loss suffered by magneticons passing through material. (In this latter case, the presence of the magnetic photon would be made explicitly manifest by its contribution to the energy loss, roughly a factor of two: see Fig. 4.) Also, as was shown by the calculations given in Appendix A, we see that the existence of magneticons will improve the agreement between theoretical and experimental values for $a_\mu$ (while not significantly affecting the agreement in the case of $a_e$). In closing, we argue that this paper gives a clear motivation for further experimental (and theoretical) study.

**ACKNOWLEDGMENTS**

I wish to thank M. K. Sullivan for numerous useful discussions, comments, and references; J. D. Bjorken for useful discussions and advice; R. Bartoldus, S. J. Brodsky, D. B. MacFarlane, M. C. Ross, and Su Dong for useful discussions and references; and G. L. Godfrey and B. N. Ratcliff for some important references.

**APPENDIX A. Feynman diagrams in QEMD**

This appendix will consider in turn further details of the QEMD calculations that are discussed in the text, namely: electron-magneticon scattering, magneticon vacuum polarization, and the magneticon-facilitated light-by-light scattering contributions to the e and μ anomalous magnetic moments. We see that electron-magneticon scattering reveals an explicit example for the difference between the one-photon and the two-photon formulations. The implications of the two-photon formulation are also discussed for the calculations for the magneticon contributions to the e and μ anomalous magnetic moments. In addition, there are implications for the stopping power calculations in Appendix B.

1. Electron-magneticon scattering

Bjorken and Drell [21] consider electron (amplitude) scattering from (the Coulomb field of) a proton, a *t*-channel process, and show that this interaction is simply a more general interaction than is electron-Coulomb scattering. The limit in which e-P scattering reduces to e-Coulomb scattering is simply that the full relativistic electron energy be much less than the proton mass. Thus, e-P scattering makes a useful reference template to use to analyze electron-magneticon scattering in the context of QEMD. Hence, we first record the Lorentz invariant matrix element, or invariant amplitude for e-P scattering [21] (again in natural units), the diagram for which is given in Fig. 3:

$$\mathcal{M}_{fi}^{(eP)} = \left[\bar{u}(e_f)\gamma^\mu u(e_i)\right]\frac{-ie^2}{q^2 + i\varepsilon}\left[\bar{u}(P_f)\gamma_\mu u(P_i)\right], \tag{A1}$$

where the arguments of the entering and exiting wavefunctions are shorthand for the relevant momenta and spins. Since we are not interested in polarization, we must average over initial spins and sum over final spins. Forming the square of the spin-averaged invariant amplitudes, then, yields:

$$\left|\overline{\mathcal{M}}_{fi}^{(eP)}\right|^2 = \frac{1}{4}\sum_{spins}\left|\bar{u}(e_f)\gamma^\mu u(e_i)\frac{e^2}{q^2+i\varepsilon}\bar{u}(P_f)\gamma_\mu u(P_i)\right|^2$$

$$= \frac{1}{4}\text{Tr}\left[\frac{(\not{p}_f + m_e)}{2m_e}\gamma^\mu\frac{(\not{p}_i + m_e)}{2m_e}\gamma^\nu\right]\text{Tr}\left[\frac{(\not{P}_f + M)}{2M}\gamma_\mu\frac{(\not{P}_i + M)}{2M}\gamma_\nu\right]\frac{e^4}{(q^2)^2}, \tag{A2}$$

where the upper case indicates proton quantities. Bjorken and Drell develop the electron trace to get:

$$\text{Tr}\left[\frac{(\not{p}_f + m_e)}{2m_e}\gamma^\mu\frac{(\not{p}_i + m_e)}{2m_e}\gamma^\nu\right] = \frac{1}{m_e^2}\left[p_f^\mu p_i^\nu + p_i^\mu p_f^\nu - g^{\mu\nu}(p_f \cdot p_i - m_e^2)\right]. \tag{A3}$$

Their final result {Ref. [21], Eq. (7.43)} for the square of the magnitude of the matrix element is:

$$\left|\overline{\mathcal{M}}_{fi}^{(eP)}\right|^2 = \frac{e^4}{2m_e^2 M^2 (q^2)^2}\left[(P_f \cdot p_f)(P_i \cdot p_i) + (P_f \cdot p_i)(P_i \cdot p_f) - m_e^2(P_f \cdot P_i) - M^2(p_f \cdot p_i) + 2M^2 m_e^2\right]. \tag{A4}$$

It is straightforward, though tedious, to show that when one lets $M \to \infty$ (the CM frame becomes the lab frame with the initial state proton at rest), one obtains from Eq. (A4) the e-Coulomb scattering cross section, as given by Eq. (9), above.

Eq. (A4) was developed assuming that the proton vertex was the Maxwell source term and the electron vertex was the Lorentz force term. (However, as we have already noted, for e-P scattering the labelling of the vertices doesn't matter; Ref. [21] shows that either way one gets the same mathematical expression, and hence only one Feynman diagram is required.) But we have indicated that the labelling of the vertices can matter in QEMD, and that one may expect different results depending upon the labelling of the vertices. The Feynman diagrams for the two electron-magneticon scattering possibilities are depicted in Fig 1. Using the above logic, it is straightforward to develop the matrix element for electron-magneticon (e-m) scattering for both cases. We first assume that the magneticon replaces the proton and is the Maxwell source, while the electron is acted on by a Lorentz force cross term, as indicated in Fig. 1a. Following this logic and the above analysis, we write for the invariant amplitude:

$$\mathcal{M}_{fi}^{(em)_a} = \left[\bar{u}(e_f)\gamma^\mu \gamma_5 u(e_i)\right] \frac{iee_m}{q^2 + i\varepsilon} \left[\bar{u}(m_f)\gamma_\mu u(m_i)\right], \tag{A5}$$

where we note that an additional minus sign is introduced with the Lorentz force cross term in the electron leg. (This calculation is just the e-Coulomb cross section calculation of Ref. [20], except here we consider a unit charged magneticon of finite mass.) The subscript a (on the amplitude superscript) denotes that we are calculating this amplitude for the diagram in Fig. 1a. We now square this amplitude and, in step-by-step analysis, obtain

$$\left|\mathcal{M}_{fi}^{(em)_a}\right|^2 = \mathcal{M}_{fi}^{(em)_a}\left(\mathcal{M}_{fi}^{(em)_a}\right)^\dagger$$

$$= \left[\bar{u}(e_f)\gamma^\mu \gamma_5 u(e_i)\right]\left[\bar{u}(m_f)\gamma^\nu u(m_i)\right]\left\{\left[\bar{u}(e_f)\gamma_\nu \gamma_5 u(e_i)\right]\left[\bar{u}(m_f)\gamma_\mu u(m_i)\right]\right\}^\dagger \frac{e^2 e_m^2}{(q^2)^2}$$

$$= \left[\bar{u}(e_f)\gamma^\mu \gamma_5 u(e_i)\right]\left[\bar{u}(m_f)\gamma^\nu u(m_i)\right]\left[\bar{u}(m_f)\gamma_\mu u(m_i)\right]^\dagger \left[\bar{u}(e_f)\gamma_\nu \gamma_5 u(e_i)\right]^\dagger \frac{e^2 e_m^2}{(q^2)^2}$$

$$= \left[\bar{u}(e_f)\gamma^\mu \gamma_5 u(e_i)\right]\left[\bar{u}(m_f)\gamma^\nu u(m_i)\right]\left[\bar{u}(m_i)\gamma_\mu u(m_f)\right]\left[\bar{u}(e_i)\gamma_\nu \gamma_5 u(e_f)\right]\frac{e^2 e_m^2}{(q^2)^2}$$

$$= \left[\bar{u}(e_f)\gamma^\mu \gamma_5 u(e_i)\right]\left[\bar{u}(e_i)\gamma^\nu \gamma_5 u(e_f)\right]\left[\bar{u}(m_f)\gamma_\nu u(m_i)\right]\left[\bar{u}(m_i)\gamma_\mu u(m_f)\right]\frac{e^2 e_m^2}{(q^2)^2}, \tag{A6}$$

where the dagger indicates the Hermitian conjugate. As with the e-P scattering, we are now in a position to form the spin averaged quantity for (em)$_a$ and sum over the final state spins (take the trace):

$$\left|\overline{\mathcal{M}}_{fi}^{(em)_a}\right|^2 = \frac{1}{4}\sum_{spins}\left|\mathcal{M}_{fi}^{(em)_a}\right|^2$$

$$= \frac{1}{4}\text{Tr}\left[\frac{(\not{p}_{e_f} + m_e)}{2m_e}\gamma^\mu \gamma_5 \frac{(\not{p}_{e_i} + m_e)}{2m_e}\gamma^\nu \gamma_5\right]\text{Tr}\left[\frac{(\not{p}_{m_f} + m_m)}{2m_m}\gamma_\mu \frac{(\not{p}_{m_i} + m_m)}{2m_m}\gamma_\nu\right]\frac{e^2 e_m^2}{(q^2)^2}. \tag{A7}$$

We see that as with e-P scattering we have the product of an electron trace and a magneticon trace. Developing the electron trace for this e-m scattering case, we obtain

$$\text{Tr}\left[\frac{(\slashed{p}_{e_f}+m_e)}{2m_e}\gamma^\mu\gamma_5\frac{(\slashed{p}_{e_i}+m_e)}{2m_e}\gamma^\nu\gamma_5\right] = \frac{1}{m_e^2}\left[p_{e_f}^\mu p_{e_i}^\nu + p_{e_i}^\mu p_{e_f}^\nu - g^{\mu\nu}(p_{e_f}\cdot p_{e_i}+m_e^2)\right], \quad (A8)$$

where we (again) see the sign reversal of the relevant mass term. Using the result for e-P scattering, Eq. (A4), it is straightforward to write the final result for this (em)$_a$ scattering case:

$$\left|\overline{\mathcal{M}}_{fi}^{(em)_a}\right|^2 = \frac{e^2 e_m^2}{2m_e^2 m_m^2 (q^2)^2}\left[(p_{m_f}\cdot p_{e_f})(p_{m_i}\cdot p_{e_i}) + (p_{m_f}\cdot p_{e_i})(p_{m_i}\cdot p_{e_f}) + m_e^2(p_{m_f}\cdot p_{m_i}) - m_m^2(p_{e_f}\cdot p_{e_i}) - 2m_m^2 m_e^2\right]$$

(A9)

At this juncture, it is interesting to look to include the QEMD case for which the electron vertex is the Maxwell source term and the proton vertex represents the Lorentz force cross term. Since the initial and final states of Figs. 1a and 1b are the same, the invariant amplitude in Eq. (A5) should be replaced by

$$\mathcal{M}_{fi}^{(em)} = \mathcal{M}_{fi}^{(em)_a} + \mathcal{M}_{fi}^{(em)_b}, \quad (A10)$$

where
$$\mathcal{M}_{fi}^{(em)_b} = \left[\bar{u}(e_f)\gamma^\mu u(e_i)\right]\frac{iee_m}{q^2+i\varepsilon}\left[\bar{u}(m_f)\gamma_5\gamma_\mu u(m_i)\right]. \quad (A11)$$

Following the (em)$_a$ analysis, we now wish to form the square of the total amplitude. That is,

$$\left|\mathcal{M}_{fi}^{(em)}\right|^2 = \left(\mathcal{M}_{fi}^{(em)_a} + \mathcal{M}_{fi}^{(em)_b}\right)\left(\mathcal{M}_{fi}^{(em)_a} + \mathcal{M}_{fi}^{(em)_b}\right)^\dagger$$

$$= \left|\mathcal{M}_{fi}^{(em)_a}\right|^2 + \mathcal{M}_{fi}^{(em)_a}\left(\mathcal{M}_{fi}^{(em)_b}\right)^\dagger + \mathcal{M}_{fi}^{(em)_b}\left(\mathcal{M}_{fi}^{(em)_a}\right)^\dagger + \left|\mathcal{M}_{fi}^{(em)_b}\right|^2. \quad (A12)$$

We continue by looking at the first interference term, for which we write

$$\mathcal{M}_{fi}^{(em)_a}\left(\mathcal{M}_{fi}^{(em)_b}\right)^\dagger = \left[\bar{u}(e_f)\gamma^\mu\gamma_5 u(e_i)\right]\left[\bar{u}(m_f)\gamma^\nu u(m_i)\right]\left\{\left[\bar{u}(e_f)\gamma_\nu u(e_i)\right]\left[\bar{u}(m_f)\gamma_5\gamma_\mu u(m_i)\right]\right\}^\dagger \frac{e^2 e_m^2}{(q^2)^2}$$

$$= \left[\bar{u}(e_f)\gamma^\mu\gamma_5 u(e_i)\right]\left[\bar{u}(m_f)\gamma^\nu u(m_i)\right]\left[\bar{u}(m_f)\gamma_5\gamma_\mu u(m_i)\right]^\dagger \left[\bar{u}(e_f)\gamma_\nu u(e_i)\right]^\dagger \frac{e^2 e_m^2}{(q^2)^2}$$

$$= \left[\bar{u}(e_f)\gamma^\mu\gamma_5 u(e_i)\right]\left[\bar{u}(m_f)\gamma^\nu u(m_i)\right]\left[\bar{u}(m_i)\gamma_5\gamma_\mu u(m_f)\right]\left[\bar{u}(e_i)\gamma_\nu u(e_f)\right]\frac{e^2 e_m^2}{(q^2)^2}$$

$$= \left[ \bar{u}(\mathrm{e_f})\gamma^\mu\gamma_5 u(\mathrm{e_i}) \right]\left[ \bar{u}(\mathrm{e_i})\gamma^\nu u(\mathrm{e_f}) \right]\left[ \bar{u}(\mathrm{m_f})\gamma_\nu u(\mathrm{m_i}) \right]\left[ \bar{u}(\mathrm{m_i})\gamma_5\gamma_\mu u(\mathrm{m_f}) \right] \frac{e^2 e_m^2}{(q^2)^2}. \quad (A13)$$

As before, we form our spin averaged quantity (summed over the final state spins) for this term:

$$\left\langle \mathcal{M}_{fi}^{(em)_a} \left( \mathcal{M}_{fi}^{(em)_b} \right)^\dagger \right\rangle = \frac{1}{4} \sum_{\text{spins}} \mathcal{M}_{fi}^{(em)_a} \left( \mathcal{M}_{fi}^{(em)_b} \right)^\dagger$$

$$= \frac{1}{4} \mathrm{Tr}\left[ \frac{(\slashed{p}_{e_f}+m_e)}{2m_e} \gamma^\mu \gamma_5 \frac{(\slashed{p}_{e_i}+m_e)}{2m_e} \gamma^\nu \right] \mathrm{Tr}\left[ \frac{(\slashed{p}_{m_f}+m_m)}{2m_m} \gamma_\nu \frac{(\slashed{p}_{m_i}+m_m)}{2m_m} \gamma_5\gamma_\mu \right] \frac{e^2 e_m^2}{(q^2)^2}, \quad (A14)$$

where $\langle \, \rangle$ indicates the averaging of the initial spins. Because the electron and magneticon are distinct and distinguishable particles, we see that unlike $e^- e^-$ scattering, this calculation does not result in one long trace for the interference terms (Here, there is no $u$-channel contribution.), but rather the product of two traces, as in e-P scattering. Each trace has terms with 2, 3, or 4 gamma matrices, each with a single $\gamma_5$. The trace theorems dictate that only those terms with 4 gamma matrices can survive. In this instance we have the relationship [21]:

$$\mathrm{Tr}\left[ \gamma_5 \slashed{a}\slashed{b}\slashed{c}\slashed{d} \right] = 4i\varepsilon_{\alpha\beta\gamma\delta} a^\alpha b^\beta c^\gamma d^\delta, \quad (A15)$$

where $\varepsilon_{\alpha\beta\gamma\delta}$ is the totally antisymmetric tensor: $\varepsilon_{\alpha\beta\gamma\delta} = 0$ if any two indices are the same, $+1$ for even permutations of $(0,1,2,3)$, and $-1$ for odd permutations. Thus, there are 24 non-zero terms, 12 with a plus sign and 12 with a minus sign. To consider Eq. (A15) in more detail, we go to the of CM frame with the initial momenta along the $z$-axis. This means that two of the four momenta (the initial state) will have null $x$ and $y$ components, with their $z$ components equal and opposite. In the final state, all momentum components of the magneticon are equal and opposite to those of the electron. When one looks in detail at these conditions on momenta, one finds that for each positive term there is an equal and opposite negative term. Thus, it follows that this interference term for electron-magneticon scattering is null. It follows, by the same logic, that the other interference term is also null.

Since the interference terms are null, we can write for e-m scattering

$$\left| \mathcal{M}_{fi}^{(em)} \right|^2 = \left| \mathcal{M}_{fi}^{(em)_a} \right|^2 + \left| \mathcal{M}_{fi}^{(em)_b} \right|^2 \quad (A16)$$

and

$$\frac{d\bar{\sigma}_{em}}{d\Omega} = \frac{d\bar{\sigma}_{em_a}}{d\Omega} + \frac{d\bar{\sigma}_{em_b}}{d\Omega}. \quad (A17)$$

For $\left| \mathcal{M}_{fi}^{(em)_b} \right|^2$, it is clear that by inspection of Eq. (A9), we can write:

$$\left|\mathcal{M}_{fi}^{(em)_b}\right|^2 = \frac{e^2 e_m^2}{2m_e^2 m_m^2 (q^2)^2}\left[(p_{m_f} \cdot p_{e_f})(p_{m_i} \cdot p_{e_i}) + (p_{m_f} \cdot p_{e_i})(p_{m_i} \cdot p_{e_f}) - m_e^2(p_{m_f} \cdot p_{m_i}) + m_m^2(p_{e_f} \cdot p_{e_i}) - 2m_m^2 m_e^2\right]$$

(A18)

for the e-m scattering process depicted in Fig. 1b. Putting Eqs. (A9 and A18) into Eq. (A16) we obtain

$$\left|\mathcal{M}_{fi}^{(em)}\right|^2 = \frac{e^2 e_m^2}{m_e^2 m_m^2 (q^2)^2}\left[(p_{m_f} \cdot p_{e_f})(p_{m_i} \cdot p_{e_i}) + (p_{m_f} \cdot p_{e_i})(p_{m_i} \cdot p_{e_f}) - 2m_m^2 m_e^2\right]. \quad \text{(A19)}$$

To compare with Eq. (9), it is now of interest to develop $(em)_a$ and $(em)_b$ scattering, which derive from Eq. (A9) and Eq. (A18). This will give the result of the magneticon's Lorentz cross force acting on the electron, and the electron's Lorentz cross force acting on the magneticon, respectively. We also look at the total scattering described by Eq. (A19). The criterion for this limit is that the electron energy $E_e = \left(m_e^2 + p_e^2\right)^{1/2} \ll m_m$. [We note that for $m_m \sim 5$ GeV/$c^2$ (the lightest magneticon allowed by extant data), this restriction still permits a rather relativistic electron, $\gamma \sim 100$ or more, say.] As before, we go to the CM frame and take the initial electron momentum to be moving in the +z direction. We assume that the electron is elastically scattered by an angle $\theta$ and choose the coordinate system such that the scattered electron momentum (as well as that of the magneticon) lies in the x-z plane. For this specific case, we write:

$$p_{e_i} = (E_{e_i}, 0, 0, p_{z_{e_i}}), \quad p_{e_f} = (E_{e_i}, p_{z_{e_i}}\sin\theta, 0, p_{z_{e_i}}\cos\theta),$$
$$p_{m_i} = (E_{m_i}, 0, 0, -p_{z_{e_i}}), \quad \text{and} \quad p_{m_f} = (E_{m_i}, -p_{z_{e_i}}\sin\theta, 0, -p_{z_{e_i}}\cos\theta),$$

(A20)

where we have used the fact that this is elastic scattering in the CM frame to simplify the terms. Also, note that the subscript z indicates the z-component of the three-momentum. We remark that from the conservation of momentum in the CM frame, we can write:

$$\beta_{m_i}^2 = \beta_{m_f}^2 = \left(\frac{\gamma_e m_e}{\gamma_m m_m}\right)^2 \beta_{e_i}^2, \quad \text{(A21)}$$

which will enable us to eliminate a number of terms that become negligible in the limit of large magneticon mass.

Using the above equations in Eq. (A9), and cancelling the $m_m^2$ factors, we can write:

$$\left|\overline{\mathcal{M}}_{fi}^{(em)_a}\right|^2 \Rightarrow \frac{e^2 e_m^2}{2m_e^2 m_m^2 (q^2)^2}\left[(E_m E_{e_f})(E_m E_{e_i}) + (E_m E_{e_i})(E_m E_{e_f}) + m_e^2(E_m E_m) - m_m^2(p_{e_f} \cdot p_{e_i}) - 2m_m^2 m_e^2\right]$$

$$= \frac{e^2 e_m^2}{2m_e^2(q^2)^2}\left[2E_{e_f}E_{e_i} - (p_{e_f} \cdot p_{e_i}) - m_e^2\right]$$
$$= \frac{8\pi^2 \alpha \alpha_m}{m_e^2(q^2)^2}\left[2E_{e_f}E_{e_i} - (p_{e_f} \cdot p_{e_i}) - m_e^2\right],$$
(A22)

which, except for the sign reversal of the $m_e^2$ term, is exactly the result of Ref. [21] for the Mott cross section. In this limit, the Mott cross section is given by [21]

$$\frac{d\bar{\sigma}_{eP}}{d\Omega} = \frac{m_e^2}{4\pi^2}\left|\mathcal{M}_{fi}^{(eP)}\right|^2.$$
(A23)

To help identify the above results with those for electron-magnetic charge scattering [Eq. (8), above], we also need for this limit

$$(q^2)^2 = 16 p_e^4 \sin^4(\theta/2).$$
(A24)

To complete this exercise, we can use the matrix element (squared) given by Eq. (A22) in Eq. (A23). Also, using Eq.(A24), we obtain, after some simplification:

$$\frac{d\bar{\sigma}_{em_a}}{d\Omega} = \frac{\alpha\alpha_m p_e^2[1+\cos(\theta)]}{8p_e^4 \sin^4(\theta/2)} = \frac{\alpha\alpha_m(1+\xi)}{2p_e^2(1-\xi)^2},$$
(A25)

where $\xi = \cos\theta$. This reproduces exactly the result for an electron scattering off of a unit of magnetic charge, Eq. (8).

We now undertake the same analysis for Eq. (A18), above, obtaining

$$\left|\overline{\mathcal{M}}_{fi}^{(em)_b}\right|^2 \Rightarrow \frac{e^2 e_m^2}{2m_e^2 m_m^2(q^2)^2}\left[(E_m \cdot p_{e_f})(E_m \cdot p_{e_i}) + (E_m \cdot p_{e_i})(E_m \cdot p_{e_f}) - m_e^2(E_m E_m) + m_m^2(p_{e_f} \cdot p_{e_i}) - 2m_m^2 m_e^2\right]$$

$$= \frac{e^2 e_m^2}{2m_e^2(q^2)^2}\left[2E_{e_f}E_{e_i} + (p_{e_f} \cdot p_{e_i}) - 3m_e^2\right]$$
$$= \frac{8\pi^2 \alpha \alpha_m}{m_e^2(q^2)^2}\left[2E_{e_f}E_{e_i} + (p_{e_f} \cdot p_{e_i}) - 3m_e^2\right],$$
(A26)

which is significantly different from Eq. (A22), above. Going through the analogous substitutions, we obtain:

$$\frac{d\bar{\sigma}_{em_b}}{d\Omega} = \frac{\alpha\alpha_m[3-\cos(\theta)]}{8p_e^2 \sin^4(\theta/2)} = \frac{\alpha\alpha_m(3-\xi)}{2p_e^2(1-\xi)^2}.$$
(A27)

One sees that both cross sections [Cf., Eq. (A25)] have the same magnitude in the forward direction ($\theta \sim 0$, or $\xi \sim 1$). However, the scattering associated with the numerator of Eq. (A25), Fig. 1a, falls to zero in the backward direction, while that associated with Fig. 1b doubles. Putting Eqs. (A46 and A27) into Eq. (A17), one obtains for the total differential cross section in the 2γ formulation:

$$\frac{d\bar{\sigma}_{em}}{d\Omega} = \frac{\alpha \alpha_m}{2 p_e^2 \sin^4(\theta/2)} = \frac{2\alpha \alpha_m}{(1-\xi)^2}. \tag{A28}$$

One can compare this result to that for e-P scattering, Eq. (9). At once, one sees two important differences: (1) the factor(s) of $\beta^2$ are absent, and (2) in the near forward direction, which accounts for the major portion of the cross section, the e-m scattering will be twice that for e-P scattering. This enhancement factor comes from the (postulated) existence of a second photon. And the missing $\beta^2$ in the denominator essentially eliminates for slow magneticons the substantial increase in cross section that characterizes slow electrically charged particles.

2. Contribution of magneticon VP loops to $a_e$ and $a_\mu$

To embark on our logical path to obtain an evaluation for $a_\ell^{QEMD}$, we refer first to Fig. 7, which is the QED Feynman diagram for the contribution to $a_\ell$ of the lowest order heavy lepton VP loop (e. g., a tau). For a reference base and a template, following the notation of Ref. [21, p. 153], i. e., natural units with $\hbar = c = 1$ and $\alpha = e^2/4\pi$ (except for the sign of $e$, which we have already discussed), we write the QED Feynman expression for the photon as it propagates from vertex 1 to vertex 4:

$$(ie\gamma_\nu)\frac{-ig^{\nu\beta}}{k_1^2}\left[(ie\gamma_\beta)\frac{i(\slashed{k}_2 + m_\tau)}{k_2^2 - m_\tau^2}(ie\gamma_\alpha)\frac{i(\slashed{k}_2 + \slashed{k}_1 + m_\tau)}{(k_2+k_1)^2 - m_\tau^2}\right]\frac{-ig^{\alpha\mu}}{k_1^2}(ie\gamma_\mu), \tag{A29}$$

where we have omitted the integral over $d^4k_2$ and its associated factors, as well as the $i\varepsilon$ terms in the denominators of the propagators. Also, we have omitted the usual $(-1)$ factor associated with closed fermion loops. In this expression, in keeping with convention, the initial vertex (1) is at the right and the final vertex (4) is at the left. The square brackets contain the VP loop portion of the expression, for which the trace will be taken to properly sum the contributions of all of the fermion components. Note that for this exercise we have routed the initial lepton momentum through the vertex at **X**, and have omitted the expressions for the fermion lines that pass through vertices 1 and 4. Also, we have routed $k_1$ through only one of the two fermion propagators that comprise the $k_2$ VP loop. As explained in Ref. [70], the mathematical expression for a given Feynman diagram is not unique; the choice of the routing of the internal circulating momenta through the loops of the Feynman network has flexibility, as long as one properly conserves momentum and energy at the vertices.

In Fig. 8 we have indicated (in parentheses) at each vertex the analogous QEMD vertex factors, for when the photon would contain a magneticon VP loop. Vertices 1 and 3 are viewed as Maxwell source vertices, while vertices 2 and 4 are viewed as Lorentz cross force vertices. (The same would be true for QED, but as we said above, it doesn't matter for QED.) As discussed above, the $\gamma_5$ factors are introduced at the Lorentz cross force "absorption" vertices. We observe that in keeping with the Maxwell source term concept, at this juncture, the photon travelling from vertex 1 to vertex 2 would be an electric photon emitted with the vertex factor $(ie\gamma_\mu)$, and the photon travelling from vertex 3 to vertex 4 would be a magnetic photon emitted with the $(-ie_m\gamma_\beta)$ factor.

For this (analogous) QEMD process, we write:

$$(-ie\gamma_\nu\gamma_5)\frac{-ig^{\nu\beta}}{k_1^2}\left[(-ie_m\gamma_\beta)\frac{i(\not{k}_2+m_m)}{k_2^2-m_m^2}(ie_m\gamma_5\gamma_\alpha)\frac{i(\not{k}_2+\not{k}_1+m_m)}{(k_2+k_1)^2-m_m^2}\right]\frac{-ig^{\alpha\mu}}{k_1^2}(ie\gamma_\mu). \quad (A30)$$

As in Eq. (A29), the square brackets indicate the VP loop portion.

It is important to note that in contrast with Fig. 2b, which has only one vertex with a $\gamma_5$ factor, the magneticon VP loop of Fig. 8 introduces a pair of $\gamma_5$ vertices into the amplitude, one to go from the electric world into the magnetic world, and a second one to return.

At this point, we make the argument that this is the only $n = 2$ magneticon VP diagram that is relevant. Basically, the argument is that the photon containing the VP loop is analogous to the *s*-channel photon in magnetic pair production (Fig. 2b). That is, only an electric photon can be emitted at vertex 1 (an electron cannot emit a magnetic photon). This photon can then create at vertex 2 either a regular lepton VP loop, e. g., a tau as in Eq. (A29), or a magneticon VP loop by a LFc term, as indicated in Eq. (A30). The magneticon VP loop in turn can only emit a magnetic photon at vertex 3, which is then absorbed at vertex 4 with a LFc interaction, also as indicated in Eq. (A30). This process cannot go in reverse from the vertex 4 because vertex 4 is on the final state electron leg, and hence is later in time than vertex 1, and photons cannot propagate backwards in time. In this context, then, vertex 1 is always a MS vertex, and vertex 4 is always a LFc term vertex. It follows, then, that Fig. 8 represents the only possible Feynman diagram for this process, even though we contemplate the existence of two photons.

Rather than actually calculate the Feynman diagram represented by Fig. 8, we intend to show that Eq. (A30) is mathematically equivalent to Eq. (A 29). To do this, we move the $\gamma_5$ factors into juxtaposition and then eliminate them from the QEMD amplitude by using Eq. (16). Pursuing this goal, if we now move the left hand (vertex 4) $\gamma_5$ factor [along with its $(-1)$ factor] along the photon line and place it in front of the vertex 3 $\gamma_\beta$ factor in the (trace) brackets, we will, in this instance, have effectively inverted the labelling of the Maxwell and Lorentz vertices.[71] (In this procedure, we are pursuing mathematics, not physics.) After this step we have:

$$(ie\gamma_\nu)\frac{-ig^{\nu\beta}}{k_1^2}\left[(ie_m\gamma_5\gamma_\beta)\frac{i(\not{k}_2+m_m)}{k_2^2-m_m^2}(ie_m\gamma_5\gamma_\alpha)\frac{i(\not{k}_2+\not{k}_1+m_m)}{(k_2+k_1)^2-m_m^2}\right]\frac{-ig^{\alpha\mu}}{k_1^2}(ie\gamma_\mu). \quad (A31)$$

We then continue to move this $\gamma_5$ to the right and form the product $(\gamma_5)^2 = 1$ with the $\gamma_5$ at vertex 2. In this process we first (commuting with the $\gamma_\beta$) have introduced an overall $(-1)$ factor to the expression (which we appropriately leave at the vertex 3 factor), and then reversed the momentum factor $\not{k}_2$ in the numerator of the fermion propagator. We record this intermediate result below:

$$(ie\gamma_\nu)\frac{-ig^{\nu\beta}}{k_1^2}\left[(-ie_m\gamma_\beta)\frac{i(-\not{k}_2+m_m)}{k_2^2-m_m^2}(ie_m\gamma_\alpha)\frac{i(\not{k}_2+\not{k}_1+m_m)}{(k_2+k_1)^2-m_m^2}\right]\frac{-ig^{\alpha\mu}}{k_1^2}(ie\gamma_\mu). \tag{A32}$$

The next step is to correct the sign of the $\not{k}_2$ (which re-establishes the conservation of momentum and energy for this expression, but also reverses the sign of $m_m$) and put in the appropriate (minus) sign for the vertex 2 factor. (These two minus 1 factors cancel out.):

$$(ie\gamma_\nu)\frac{-ig^{\nu\beta}}{k_1^2}\left[(-ie_m\gamma_\beta)\frac{i(\not{k}_2-m_m)}{k_2^2-m_m^2}(-ie_m\gamma_\alpha)\frac{i(\not{k}_2+\not{k}_1+m_m)}{(k_2+k_1)^2-m_m^2}\right]\frac{-ig^{\alpha\mu}}{k_1^2}(ie\gamma_\mu). \tag{A33}$$

One sees that we have achieved a more detailed agreement of the QEMD expression, Eq. (A33), with the original QED expression, Eq. (A29). In fact, we have an exact duplication of Eq. (A29) except for the minus signs of the charges (which we have already explained) and the sign of $m_m$ in the numerator of the left hand magneticon propagator, which discrepancy we now address.

From a qualitative point of view, it is obvious that the major contributions to the integral of the Eq. (A33) will be associated with the regions in momentum space for which the propagating fermions are near their mass shell, i. e. $k^2 \sim m_m^2$, a condition pertaining to the denominators of the propagators. This means that we expect the mass of the fermion to make its major contribution to the integrals via the denominators of the fermion propagators. Turning to the mathematics, in evaluating the Feynman integrals, it is frequent practice to elevate these propagator denominators into the arguments of exponentials by the identity [21, Eq. (8.12)]:

$$\frac{i}{\not{k}-m+i\varepsilon} = \frac{i(\not{k}+m)}{k^2-m^2+i\varepsilon} = (\not{k}+m)\int_0^\infty dz\exp[iz(k^2-m^2+i\varepsilon)]. \tag{A34}$$

This conversion, which assists in the mathematical evaluation of the Feynman diagrams, also makes it easier to follow, through the mathematical steps, the contribution of the mass factor in the numerator versus that of the mass factor in the denominator. Without going through the rather intricate details of the mathematics, which can be found in Ref. [21], it was shown there that the contributions of the mass terms deriving from the numerators of the propagators of the VP loop actually vanish. The significance of this fact is that the sign reversal of $m_m$ in Eq. (A33) will not affect the evaluation of the magneticon VP loop contribution to $a_\ell$. That is, there is no surviving difference between the QED expression, Eq. (A29), and the QEMD expression, Eq. (A33). Hence, the VP loop of the magneticon contribution to $a_\ell$ can be evaluated just as though it were a heavy (electric) fermion of mass $m_m$.[72]

To proceed with this evaluation of the magneticon $a_\ell^{\text{QEMD}}$ contribution to the theoretical values for $a_e$ (and $a_\mu$, below), then, one can simply use the formulae collected by Knecht [54], who gives the theoretical expansion[73] for QED calculations of $a_\ell^{\text{QED}}$ as follows:

$$a_\ell^{\text{QED}} = \sum_{n \geq 1} A_n \left(\frac{\alpha}{\pi}\right)^n + \sum_{n \geq 2} B_n(\ell, \ell') \left(\frac{\alpha}{\pi}\right)^n, \tag{A35}$$

where the $A_n$ gives the contributions associated with $\ell$ alone, and $B_n(\ell, \ell')$ gives the contributions by $\ell'$ (through the Feynman graphs containing closed $\ell'$ loops) to $a_\ell^{\text{QED}}$, and $n$ indicates the number of integration loops in the diagram. For this analysis, it is the $B_n(\ell, \ell')$ coefficients that are of interest. In this formulation, as was argued above, all that is required to evaluate the magneticon VP loop contribution is the ratio of the mass of the magneticon to that of the lepton in question. For $n > 2$, more than one (non $\ell$) lepton may make VP contributions, for which the notation $B_n(\ell; \ell', \ell'')$ is used. Though these terms are too small to be relevant for this analysis, the modifications required for this extension are straightforward.

Using the above analysis, we can now evaluate the magnitude of the magneticon VP contribution by using the formula given by Knecht for $B_2(\ell, \ell')$:

$$\begin{aligned} B_2(\ell, \ell') = &\frac{1}{45}\left(\frac{m_\ell}{m_{\ell'}}\right)^2 + \frac{1}{70}\left(\frac{m_\ell}{m_{\ell'}}\right)^4 \ln\left(\frac{m_\ell}{m_{\ell'}}\right) + \frac{9}{19600}\left(\frac{m_\ell}{m_{\ell'}}\right)^4 \\ &+ \frac{4}{315}\left(\frac{m_\ell}{m_{\ell'}}\right)^6 \ln\left(\frac{m_\ell}{m_{\ell'}}\right) - \frac{131}{99225}\left(\frac{m_\ell}{m_{\ell'}}\right)^6 + O\left[\left(\frac{m_\ell}{m_{\ell'}}\right)^8 \ln\left(\frac{m_\ell}{m_{\ell'}}\right)\right]. \end{aligned} \tag{A36}$$

It is clear that the factors of $\left(\frac{m_\ell}{m_{\ell'}}\right)$ raised to powers of $n \geq 2$ and the factor $\left(\frac{\alpha}{\pi}\right)^2$ associated with the various terms of this expression for $B_2$ will significantly reduce the magnitude of the VP contribution of magneticon loops to both $a_e$ and $a_\mu$. In Eq. (A36) there are five specified terms in this expansion of $B_2$, as well as an indication of the order of magnitude of the next term in the expansion. The leading term, is the largest for both the electron and muon. In both cases it is a straightforward calculation to show that the higher order terms are three or more orders of magnitude smaller than the leading term, and hence, for our present exercise, may be neglected. For the electron, using the values for $m_e$, as given by the PDG and $m_m = 5.1$ GeV/$c^2$, we obtain $B_2(e,m) = 2.23 \times 10^{-10}$, which gives the magneticon contribution $a_e^{\text{QEMD}} = B_2(e,m) \times (\alpha/\pi)^2 = 1.2 \times 10^{-15}$, which has been included in Table I. The same logic was also used to calculate $a_\mu^{\text{QEMD}} = B_2(\mu,m) \times (\alpha/\pi)^2 = 5.14 \times 10^{-11}$, which was added to Table II.

Clearly, this calculated value for $a_e^{\text{QEMD}}$ [and also, that for $a_e^{\text{L}\times\text{L}}(e;m)$, see discussion below] is negligible with respect to the present overall error level of $a_e^{\text{QED}}$. Therefore, as done in Ref. [53], it remains legitimate to write the expansion:

$$a_e^{\text{QED}} = \sum_{n=1}^{5} C_e^{(2n)} \left(\frac{\alpha}{\pi}\right)^n, \tag{A37}$$

where the coefficients (with uncertainties, when relevant) are: $C_e^{(2)} = 0.5$, $C_e^{(4)} = -0.328\,478\,444\,00$, $C_e^{(6)} = 1.181\,234\,017$, $C_e^{(8)} = -1.9144(35)$, and $C_e^{(10)} = 0.0(4.6)$. It turns out that the uncertainty of $C_e^{(10)}$ dominates the uncertainty of Eq. (A8), and the terms for $n > 5$ are not yet relevant. Ref. [53] also gives the best values (with associated uncertainties) for $a_e^{\text{wk}}$, and $a_e^{\text{had}}$, which are $0.029\,73(52)\times 10^{-12}$ and $1.685(22)\times 10^{-12}$, respectively.

3. Magneticon-facilitated light-by-light scattering contribution to $a_e$ and $a_\mu$

To estimate the 3-loop light-by-light contribution to $a_\ell$, we consider the appropriate Feynman diagrams (there are 6 of them), three of which are depicted in Fig. 9. The three additional diagrams arise when one reverses the direction of circulation of the fermion loop. In Fig. 10 we show in more detail one of the diagrams in Fig. 9. As in Figs. 7 and 8, the difference between the QED and QEMD lies in the vertex factors The vertex factors in Fig. 10 are those for a standard QED heavy lepton, and those in the QEMD analogue diagram are those for a magneticon, as explained above. In the case of $B_2(\ell,\ell')$, for the magneticon VP loops, we showed that one can calculate the QEMD magneticon contribution using the QED formulae. A similar but more complicated argument can be made for the QEMD-QED relationship in the case of the magneticon loop in the contribution to $a_\ell^{\text{L}\times\text{L}}(\ell;m)$. As we did in the case of $B_2(\ell,\ell')$, we first write out the QED expression for the $a_\ell^{\text{L}\times\text{L}}(\ell;\tau)$ diagram as shown in Fig. 10. Schematically, we have:

$$\text{QED expression} = [\text{closed loop}]_E \{\text{lepton line}\}_E, \tag{A38}$$

where the subscript E denotes a QED formulation In more detail, we have [Again, we omit the (−1) and integration factors associated with the closed fermion loop.]:

$$[\ ]_E = \left[(ie\gamma_\mu)\frac{i(\slashed{k}_1 + m_\tau)}{k_1^2 - m_\tau^2}(ie\gamma_\beta)\frac{i(\slashed{k}_1 - \slashed{k}_3 + m_\tau)}{(k_1-k_3)^2 - m_\tau^2}(ie\gamma_\alpha)\frac{i(\slashed{k}_1 - \slashed{k}_2 + m_\tau)}{(k_1-k_2)^2 - m_\tau^2}(ie\gamma_\nu)\frac{i(\slashed{k}_1 + m_\tau)}{k_1^2 - m_\tau^2}\right] \tag{A39}$$

and

$$\{\text{lepton line}\}_E = \left\{(ie\gamma^\nu)\frac{i(\slashed{p} - \slashed{k}_2 + m_\ell)}{(p-k_2)^2 - m_\ell^2}(ie\gamma^\alpha)\frac{i(\slashed{p} - \slashed{k}_3 + m_\ell)}{(p-k_3)^2 - m_\ell^2}(ie\gamma^\beta)\right\}. \tag{A40}$$

As before, we now write the analogous QEMD expressions:

$$[\ ]_M = \left[ (ie_m \gamma_5 \gamma_\mu) \frac{i(\slashed{k}_1 + m_m)}{k_1^2 - m_m^2} (-ie_m \gamma_\beta) \frac{i(\slashed{k}_1 - \slashed{k}_3 + m_m)}{(k_1 - k_3)^2 - m_m^2} (-ie_m \gamma_\alpha) \frac{i(\slashed{k}_1 - \slashed{k}_2 + m_m)}{(k_1 - k_2)^2 - m_m^2} (-ie_m \gamma_\nu) \frac{i(\slashed{k}_1 + m_m)}{k_1^2 - m_m^2} \right]$$

(A41)

and

$$\{\text{lepton line}\}_M = \left\{ (-ie\gamma^\nu \gamma_5) \frac{i(\slashed{p} - \slashed{k}_2 + m_\ell)}{(p - k_2)^2 - m_\ell^2} (-ie\gamma^\alpha \gamma_5) \frac{i(\slashed{p} - \slashed{k}_3 + m_\ell)}{(p - k_3)^2 - m_\ell^2} (-ie\gamma^\beta \gamma_5) \right\},$$

(A42)

where the subscript M denotes the QEMD formulation. The $\gamma_5$s are included in the vertex factors that we associate with the Lorentz force cross terms. It can also be seen that we have not included the external lepton legs and the photon propagators, which do not enter into our manipulations. The superscripts and subscripts indicate which gammas connect which vertices. The momentum factors in the fermion propagators are consistent with these notations. We remark that we view vertex 7 in the Feynman diagram in Fig. 11 as part of a *t*-channel diagram, analogous to the magneticon vertex in Fig. 2b. Following this logic, the lepton line vertices (1, 2, 3) are analogous to the lepton vertex in Fig. 2a.

In order to eliminate the $\gamma_5$s, as with the magneticon VP diagram, we start with the lepton line of Eq. (A42), and move the $\gamma_5$s (and their minus signs) along their respective photon lines into the magneticon loop bracket $[\ ]_M$. [We note again that this step is equivalent to reversing the labelling of the pair of vertices, and argue, as before, that for closed interior magneticon loops that do not connect (through a single photon) to an external magneticon leg, the vertex labelling doesn't matter.] This step gives us:

$$\{\text{lepton line}\}_M = \left\{ (ie\gamma^\nu) \frac{i(\slashed{p} - \slashed{k}_2 + m_\ell)}{(p - k_2)^2 - m_\ell^2} (ie\gamma^\alpha) \frac{i(\slashed{p} - \slashed{k}_3 + m_\ell)}{(p - k_3)^2 - m_\ell^2} (ie\gamma^\beta) \right\}$$

(A43)

and

$$[\ ]_M = \left[ (ie_m \gamma_5 \gamma_\mu) \frac{i(\slashed{k}_1 + m_m)}{k_1^2 - m_m^2} (ie_m \gamma_5 \gamma_\beta) \frac{i(\slashed{k}_1 - \slashed{k}_3 + m_m)}{(k_1 - k_3)^2 - m_m^2} (ie_m \gamma_5 \gamma_\alpha) \frac{i(\slashed{k}_1 - \slashed{k}_2 + m_m)}{(k_1 - k_2)^2 - m_m^2} (ie_m \gamma_5 \gamma_\nu) \frac{i(\slashed{k}_1 + m_m)}{k_1^2 - m_m^2} \right].$$

(A44)

We see that the $\{\text{lepton line}\}_M$ is now in complete agreement with the $\{\text{lepton line}\}_E$.

At this point, we are in a position to eliminate the $\gamma_5$s in the magneticon loop. As before, combining the $\gamma_5$s in pairs, we get a pair of sign reversals (through $\gamma$ anticommutation), which we can then use to reinstate the momentum arguments in the affected fermion propagators. This step leaves us with:

$$[\ ]_M = \left[(ie_m\gamma_\mu)\frac{i(\slashed{k}_1 - m_m)}{k_1^2 - m_m^2}(ie_m\gamma_\beta)\frac{i(\slashed{k}_1 - \slashed{k}_3 + m_m)}{(k_1 - k_3)^2 - m_m^2}(ie_m\gamma_\alpha)\frac{i(\slashed{k}_1 - \slashed{k}_2 - m_m)}{(k_1 - k_2)^2 - m_m^2}(ie_m\gamma_\nu)\frac{i(\slashed{k}_1 + m_m)}{k_1^2 - m_m^2}\right]. \quad (A45)$$

We can now change the signs in the four vertex factors $[(-1)^4 = 1]$ to obtain our final expression:

$$[\ ]_M = \left[(-ie_m\gamma_\mu)\frac{i(\slashed{k}_1 - m_m)}{k_1^2 - m_m^2}(-ie_m\gamma_\beta)\frac{i(\slashed{k}_1 - \slashed{k}_3 + m_m)}{(k_1 - k_3)^2 - m_m^2}(-ie_m\gamma_\alpha)\frac{i(\slashed{k}_1 - \slashed{k}_2 - m_m)}{(k_1 - k_2)^2 - m_m^2}(-ie_m\gamma_\nu)\frac{i(\slashed{k}_1 + m_m)}{k_1^2 - m_m^2}\right]. \quad (A46)$$

We have in Eq. (A46) the same discrepancy as we had in Eq. (A33), namely the reversal in the sign of $m_m$ in the numerator of two of the magneticon propagators. We conjecture that as with Eq. (A33), these mass terms with their sign reversals do not enter into the final calculation for this contribution to $a_e$ or $a_\mu$. This conjecture, of course, applies to all six of the magneticon light-by-light diagrams represented in Fig. 9. In this case, the use of the Knecht formulae for $a_\ell$ will give proper results by treating the magneticon as an electric fermion with the magneticon mass. We see that this case gives a contribution too small to be relevant to the present state of the art in either experiment or theory. (Cf., Table I and Table II.) Should this conjecture not be fully valid, we argue that the Knecht formulae can still be used to give a reasonable estimate, and that these contributions are still be too small to be relevant.

To carry out this program, for the electron, we write

$$B_3^{L\times L}(e;m) = \left(\frac{m_e}{m_m}\right)^2\left[\frac{3}{2}\zeta(3) - \frac{19}{16}\right] + \left(\frac{m_e}{m_m}\right)^4\left[-\frac{161}{810}\ln^2\left(\frac{m_m}{m_e}\right) - \frac{16189}{48600}\ln\left(\frac{m_m}{m_e}\right) + C\right] + O\left[\left(\frac{m_e}{m_m}\right)^6\right], \quad (A47)$$

where $C = \frac{13}{18}\zeta(3) - \frac{161}{9720}\pi^2 - \frac{831931}{972000}$. N. B., $B_3^{L\times L}(e;m)$ is a component of $B_3(\ell,\ell')$ of Eq. (A35).

(We note that for the electron the magnitude of the first term of this expansion exceeds that of the second by a factor of $> 10^6$.) Thus, we obtain a calculation/estimate for the lowest order magneticon light-by-light contribution to $a_e$:

$$a_e^{L\times L}(e;m) = B_3^{L\times L}(e;m)\left(\frac{\alpha}{\pi}\right)^3 = 7.7\times 10^{-17}, \quad (A48)$$

which has been included in Table I as the $1\gamma$ result. Similarly, for the muon, we obtain

$$a_\mu^{L\times L}(\mu;m) = B_3^{L\times L}(\mu;m)\left(\frac{\alpha}{\pi}\right)^3 = 3.3\times 10^{-12}, \quad (A49)$$

which has been included in Table II, also as the 1γ result. Unlike the case for $a_\ell^{\text{QED}}(\ell,\text{m})$ where we argued that the 2γ result was the same as the 1γ result, we argue that for $a_\ell^{\text{L}\times\text{L}}(\ell,\text{m})$ all four of the internal photons can be either electric or magnetic. Thus, this would augment the 2γ result by a factor of $\sim 2^4$, a factor of two for each photon.[74] However, as we argued above (see endnote [59]) that the internal magneton loop must have at least one LFc vertex. This means that one of the 16 additional diagrams will be eliminated from the multiplier, i., e., one makes the approximation:

$$a_\ell^{\text{L}\times\text{L}}(\ell,\text{m})_{2\gamma} = 15\ a_\ell^{\text{L}\times\text{L}}(\ell,\text{m})_{1\gamma}. \tag{A50}$$

The appropriate numbers have been added to Tables I and II.

**APPENDIX B. Energy loss for magneticons passing through matter**

In order to develop a basis for understanding the 1γ and 2γ formulations for *dE/dx* for magneticons passing through matter, we first review the energy loss of electrically charged particles. Jackson [15, Ch. 13],[75] following Bohr [71], begins his discussion using classical equations for the transfer of energy from an electrically charged projectile to a particle (electron) at rest. This energy transfer depends upon the velocity *v* and charge *ze* of the projectile, the charge of the electron, and the impact parameter *b* characterizing the "collision." To find the energy loss of the projectile traversing an absorbing material, he then integrates over the effective range of *b*, where Δ*E(b)* is the energy transfer, estimated classically, for a single collision. He obtains the result (in Gaussian units):

$$\frac{dE_c}{dx} = 2\pi NZ \int \Delta E(b)\,db \simeq 4\pi NZ \frac{z^2 e^4}{m_e v^2} \int_{b_{\min}}^{b_{\max}} \frac{b\,db}{b^2} = 4\pi NZ \frac{z^2 e^4}{m_e v^2} \ln B\ , \tag{B1}$$

where *N* is the number of atoms/unit volume, *Z* is the number of electrons/atom, $m_e$ is electron mass, and $B = b_{\max}/b_{\min}$, where $b_{\max}$ ($b_{\min}$) is the largest (smallest) effective impact parameter. The subscript "c" on $E_c$ indicates that this is a classical derivation. As Jackson explains, this result is a reasonably accurate approximation to Bethe's result, but it can be improved by bringing in quantum considerations. Since the leading physics multiplier in front of the integral is the same for both this classical derivation and the quantum derivation, as given by Bethe [72], the main focus for improving this derivation is in the determination of the quantities $b_{\min}$ and $b_{\max}$ in the argument of the logarithm.

Jackson then gives Bethe's formula for *dE/dx* for (electrically) charged particles traversing matter:

$$\frac{dE_q}{dx} = 4\pi NZ \frac{z^2 e^4}{m_e v^2} \left[ \ln\left(\frac{2\gamma^2 m_e v^2}{\hbar \langle \omega \rangle}\right) - \frac{v^2}{c^2} \right] \text{ ergs/cm}, \tag{B2}$$

where the subscript "q" on $E_q$ indicates the use of quantum considerations, and $v$ and $\gamma = (1 - \beta^2)^{-\frac{1}{2}}$ characterize the incident projectile. The average frequency $\langle \omega \rangle$ is a log geometric mean defined by the equation

$$Z \ln \langle \omega \rangle = \sum_j f_j \ln \omega_j, \tag{B3}$$

where $f_j$ is the strength of the $j^{th}$ oscillator, and the index $j$ is over the $Z$ electrons in the (target) atom. It is useful to note that the quantity $\hbar \langle \omega \rangle$ is essentially the same as $I$ given in Table VI of Ref. [22]. As discussed by Fano [39], the integral used to derive Eq. (B2) is taken over an appropriate range of $Q$, the lab recoil (kinetic) energy of the electron. The integral over $Q$, of course, is characterized by a maximum and minimum value, which lead to the argument of the logarithm in Eq. (B2).

In order to make the expression for $dE/dx$ more useful, we rewrite Eq. (B2):

$$\frac{dE_q(ze)}{dx} = \frac{4\pi N_A Z z^2 \alpha^2 (\hbar c)^2}{A m_e v^2} \left[ \ln \left( \frac{2\gamma^2 m_e v^2}{I} \right) - \frac{v^2}{c^2} \right] \text{MeV cm}^2/\text{g}, \tag{B4}$$

where $N_A$ is Avagadro's number (per g-mole), $A$ is the atomic weight (of a pure element) in g/g-mole and $\hbar \langle \omega \rangle$ in the argument of the logarithm has been replaced with $I$, mentioned above. By a suitable choice of units for $\hbar$, $c$, and $m_e$, the result can be obtained in MeV-cm$^2$/g. Eq. (B4) is a $1\gamma$ $dE/dx$ formulation for a projectile of charge ($ze$) as discussed above in Sec. IV of the text. Here, $x$, of course, carries the units g/cm$^2$. There are additional corrections to $dE/dx$, such as an inner shell correction and density effects [39], but these are relatively small.

To help us derive a $dE/dx$ formula for magneticons, we first develop another stopping power formulation (for electrically charged particles) based upon the cross section for an electron (Coulomb) scattering off of a proton at rest, Eq. (9). (See Fig. 3.) The purpose of this effort is to evaluate the suitability of this approach to stopping power calculations for protons and by implication for magneticons: the proper result for electrically charged particles (e. g., protons) is well known, and thus, a successful result using this formulation for protons means that we can then use the em scattering cross sections developed in Appendix A in a similar way to find the stopping power formula for magneticons. This proposed formulation will be most accurate for larger energy transfers (since the available CM energy is large compared to the electron binding energy to the nucleus). And since we are using a quantum mechanical derivation, we do not have to address the question of the appropriate value for $b_{\min}$ associated with the semi-classical derivation of Bohr. However, without correction or adjustment, it will yield a poor approximation at small energy transfers, since in this region the binding energy of the electrons becomes significant. This latter problem will be accounted for by using a suitable approximation for the scattering cutoff in the forward direction, for which the electron binding energy effectively quenches the scattering. This is a "soft" quenching, since different electrons have different binding energies, which is taken into account by the use of the parameter $I$, mentioned above.

There are three frames that are relevant for this derivation. The first is a lab frame (Frame I), in which the electron is at rest and the initial proton 3-momentum is $p$. This is the frame in which we wish to obtain the $dE/dx$ of the proton. The second frame (Frame II) is the CM frame for this (elastic) scattering problem, in which the electron is not bound. In the CM frame the proton 3-momentum is given by

$$p' = \frac{m_e p}{E'}, \tag{B5}$$

where $E'$, the total CM energy, is

$$E' = \sqrt{2 m_e E_P + M^2 + m_e^2}, \tag{B6}$$

and $E_P$ is the fourth component of the initial proton 4-momentum in Frame I. (Here, we use the relativistic equations of Jackson [15]. Also, following Jackson, we use primes to indicate the CM variables, and maintain $c = 1$ to simplify the equations.) In the CM frame we have $p'_e = -p'$, which equality enables the determination of the magnitude of the electron momentum in the third frame (Frame III), in which frame the electron is the projectile and the proton is at rest. We use double primes on the variables in Frame III.

Using the above CM relationships, one obtains

$$\gamma''_e \beta''_e = \gamma_P \beta_P. \tag{B7}$$

That is, in this formulation, the relativistic factors for the electron in Frame III are identical to those for the proton in the Frame I. We note also that since $\sigma$ is transverse to the direction of motion, it is invariant with respect to frame. Thus, taking care to use the appropriate variables, Eq. (9) needs no modification [except to include the implied $(\hbar c)^2$ factor] to be used in Frame III. Consequently, as an approximation for the energy loss of a proton projectile in Frame I, we can write

$$\frac{dE_P}{dx} = \frac{N_A Z}{A} \int \frac{d\bar{\sigma}_{eP}}{d\Omega''} T_e \, d\Omega'', \tag{B8}$$

where the integral is over the solid angle in Frame III (in which $d\bar{\sigma}_{eP}$ is calculated), and $T_e$ is the kinetic recoil energy of the electron in Frame I. The formula for $T_e$ is [15, Eq. (12.55)]:

$$T_e = \frac{m_e p^2 (1 - \cos\theta')}{2 m_e E_P + M^2 + m_e^2}, \tag{B9}$$

where $p$ is the magnitude of the initial proton 3-momentum in the Frame I. For later use, we note that $T_e$ is closely related to the (final state) electron recoil parameter $Q$ used by Fano [39]. In fact, the relationship is linear (actually a near identity) in the intermediate range of $Q$, but is in need of some modifications near $Q_{min}$ and $Q_{max}$.

Putting Eqs. (B9) and (9) into Eq. (B8), and including the $(\hbar c)^2$ for dimensional purposes, we obtain

$$\frac{dE_{\rm P}}{dx} = \frac{N_{\rm A} Z(\hbar c)^2}{A} \int \frac{\alpha^2 \left[2 - \beta_{\rm e}''^2 (1-\cos\theta'')\right]}{2 m_{\rm e} v^2 (1-\cos\theta'')^2} \frac{M^2(1-\cos\theta')}{2 m_{\rm e} E_{\rm P} + M^2 + m_{\rm e}^2} d\Omega''. \tag{B10}$$

We now observe that for the velocities relevant to this study ($\gamma < 5$, say), with a relative error of less than $5\times 10^{-3}$, we can set

$$\frac{M^2}{2 m_{\rm e} E_{\rm P} + M^2 + m_{\rm e}^2} = 1. \tag{B11}$$

But to properly carry out the indicated integration, we need to obtain $\theta'$ in terms of $\theta''$. For this purpose, we use the relationship

$$\tan\theta'' = \frac{\sin\theta''}{\cos\theta''} = \frac{E'\sin\theta'}{(\gamma m_{\rm e} + M)(\cos\theta' + \alpha)} \simeq \frac{\sin\theta'}{(\cos\theta' + \alpha)}, \tag{B12}$$

where (for the conditions for this study) the final equality is good to better than $10^{-6}$. Here, the ratio $\alpha$ is a general scattering parameter. For elastic scattering, $\alpha$ becomes $\alpha_{\rm es}$, which is given by

$$\alpha_{\rm es} = \frac{m_{\rm e}}{M} \frac{(M\gamma + m_{\rm e})}{(\gamma m_{\rm e} + M)}. \tag{B13}$$

We note that for this study that $\alpha_{\rm es} \sim 2.5\times 10^{-3}$ for protons and $\sim 5\times 10^{-4}$ for magneticons (of mass $\sim 5$ GeV/$c^2$).

Expanding Eq. (B12) about $\theta'' = 0$, we find that $\theta'' \simeq (1 - \alpha_{\rm es})\theta'$. (At $\theta'$, $\theta'' \to \pi$, $\theta'$ and $\theta''$ again converge.) Thus, if we set

$$(1-\cos\vartheta') \equiv \eta' \Rightarrow (1-\cos\vartheta'') \equiv \eta'', \tag{B14}$$

the largest relative error incurred in the range of integration is $\sim \alpha_{\rm es}$, which we shall ignore. Substituting Eq. (B14) into Eq. (B10), applying Eq. (B11), we write

$$\frac{dE_{\rm P}}{dx} = \frac{\pi N_{\rm A} Z(\hbar c)^2 \alpha^2}{A m_{\rm e} v^2} \int_{\eta''_{\rm min}}^{\eta''_{\rm max}} \frac{\left[2 - \beta^2 \eta''\right]}{\eta''} d\eta'', \tag{B15}$$

where we have integrated $d\Omega'' = d\eta'' d\phi''$ over $0 \leq \phi'' \leq 2\pi$. We observe that $\eta''_{\rm min}$, corresponds to the effective quenching limit for small deflections. Carrying out the $\eta''$ integration, we obtain

$$\frac{dE_P}{dx} = \frac{2\pi N_A Z (\hbar c)^2 \alpha^2}{A m_e v^2} \left[ \ln\left(\frac{\eta''_{max}}{\eta''_{min}}\right) - \beta^2 \right]. \tag{B16}$$

Comparing Eq. (B16) to Eq. (B4), we see that the functional form of Eq. (B16) compares well with that of Eq. (B4). That is, the leading physics factors are identical (to a factor of 2) and the main contribution to the final result is the logarithm of a large number that involves a ratio of the limits of the integration. In the square bracket, there is also an additional term, $\beta^2$, for which the integration does not entail the divergence of $d\bar{\sigma}$ in the forward direction. Hence, we have used the nominal values of $\eta''_{min}$ and $\eta''_{max}$ (0 and 2) for its evaluation. This term is half of the accepted value in the Bethe formulation, and is known to be associated with relativistic corrections. It makes a minor contribution to the final result (in the range of a few percent).

We now have the question of the proper values for $\eta''_{min}$ and $\eta''_{max}$ as the limits for the log term in the main integration, which is nominally divergent. Since the proper end point values of these parameters are obscured by the free electron modelling in this formulation, we turn to Fano [39] for guidance, which leads us to surmise that it is appropriate to set

$$\left(\frac{\eta''_{max}}{\eta''_{min}}\right) = \left(\frac{2\gamma^2 m_e v^2}{I}\right)^2. \tag{B17}$$

Putting Eq. (B17), which compensates for some of the deficiencies of the free electron modelling, into Eq. (B16), we obtain

$$\frac{dE_P}{dx} = \frac{4\pi N_A Z (\hbar c)^2 \alpha^2}{A m_e v^2} \left[ \ln\left(\frac{2\gamma^2 m_e v^2}{I}\right) - \frac{\beta^2}{2} \right]. \tag{B18}$$

The only difference now between Eq. (B18) and the Bethe result, Eq. (B2), is the magnitude of the $\beta^2$ term. As mentioned above, its contribution is relatively small, and results from the deficiency of our mathematical modelling, which uses free electrons instead of bound electrons.

For the purpose of a direct comparison of these $dE/dx$ formulations and as a numerical check, using Eq. (B18) we shall make a $dE/dx$ calculation for protons (of mass $M$) penetrating graphite (which we use to represent plastic scintillator, which by weight is mostly carbon: the fractional weight factor for carbon in plastic scintillator is 0.915 [73]; the remainder is almost all hydrogen.). This calculation, then, is to be compared to the stopping power calculation (for the same conditions, the minimum ionization point of protons on graphite) found on the NIST web site [73].[76] The NIST minimum, which is 1.746 MeV-cm$^2$/g, is found at $T_P$ = 3 GeV. For this point, $\gamma_P$ = 4.197 and $\beta_P$ = 0.9712. (This choice is also relevant our analysis of the magneticon $dE/dx$ in the FQS, because the gamma factor for a 5 GeV/$c^2$ magneticon produced by a 14.5 GeV beam will be reasonably close to minimum ionizing.) At this point, Eq. (B18) gives $dE_P/dx$ = 1.92 MeV-cm$^2$/g, which is ~10% above the NIST value. If we were to use the uncorrected Bethe formula, Eq. (B4), we would get $dE_P/dx = 1.8434$ MeV-cm$^2$/g, which is within 6% of the NIST value. The important missing correction here is the density effect, which disappears for lower values of the $\gamma$ pertaining to the projectile.

Having verified the validity of this formulation using the relativistic cross section of an electron scattering off of a proton [Eq. (9)] to obtain a useful approximation for $dE_P/dx$, we now retrace the above steps to obtain a formula for the $dE_g/dx$ for magneticons of charge $g$. As shown in Appendix A, there are two cross sections to be considered; one with the exchange of a magnetic photon, Eq. (A25), (Fig. 1a) and one with the exchange of an electric photon, Eq. (A27), (Fig. 1b).

For the first case we have:

$$\frac{dE_{g_a}}{dx} = \frac{2\pi N_A Z \alpha_m \alpha (\hbar c)^2}{A m_e c^2} \left[ \ln\left(\frac{\eta''_{max}}{\eta''_{min}}\right) - 1 \right],  \tag{B19}$$

which is the same form as Eq. (B4), but with the $v^2$ factor in the denominator replaced by $c^2$. (As before, $Z$ is the number of electrons/atom.) Similarly, for the second Feynman diagram, we have;

$$\frac{dE_{g_b}}{dx} = \frac{2\pi N_A Z \alpha_m \alpha (\hbar c)^2}{A m_e c^2} \left[ \ln\left(\frac{\eta''_{max}}{\eta''_{min}}\right) + 1 \right],  \tag{B20}$$

which is somewhat larger than Eq. (B19). Applying Eq. (B17) (which we tacitly assume to also be relevant to magneticon stopping power) to Eq. (B19), we obtain

$$\frac{dE_{em}^{1\gamma}}{dx} \equiv \frac{dE_{em_a}}{dx} = \frac{4\pi N_A Z \alpha_m \alpha (\hbar c)^2}{A m_e c^2} \left[ \ln\left(\frac{2\gamma^2 m_e v^2}{I_m}\right) - \frac{1}{2} \right] \text{ MeV cm}^2/\text{g},  \tag{B21}$$

where, following Ahlen [22], we have replaced $I$ with $I_m$ to acknowledge that this quantity may differ because of different electric and magnetic interactions of the projectile with the atomic electrons. We see here that the differences between Eq.(B4) and Eq. (B21) are the use of $\alpha_m$ to represent the magneticon charge $g$, and the elimination of the velocity factor from the denominator of the physics multiplier. Eq. (B21) is consistent with the results of Ahlen [22, 41],[77] and we shall use it as the basis for the magneticon 1$\gamma$ (re)analysis of the FQS data. As indicated by our notation, we shall use Eq. (B21) as our 1$\gamma$ formulation to study the $dE/dx$ of magneticons in the FQS (re)analysis (and, as suggested by Ahlen, assume that $I = I_m$). The main concurrence here is that near the minimum ionizing point, in this formulation, magneticons lose energy at about the same rate as do protons, however, as mentioned above, it is important to note the fact that there is no low-velocity enhancement for magnetic $dE/dx$, as has been discussed by Ahlen [22], and others.

Summing Eqs. (B19 and 20) and making the same conversions, yield the result for the 2$\gamma$ formulation for magneticons:

$$\frac{dE_{em}^{2\gamma}}{dx} = \frac{8\pi N_A Z \alpha_m \alpha (\hbar c)^2}{A m_e c^2} \ln\left(\frac{2\gamma^2 m_e v^2}{I_m}\right) \text{ MeV cm}^2/\text{g}. \tag{B22}$$

Again, we have no low-velocity enhancement, but the energy loss near the ionization minimum is about twice that for protons with the same velocity. We shall use Eq. (B22) for our 2γ formulation for the study of magneticon energy loss in the FQS.

**APPENDIX C. Solid angle and expected back-to-back pair events in the FQS**

In this appendix we will calculate the number of μ, e, and m pair events that one expects to be recorded in the solid angle of FQS apparatus [28], which equals $4\pi/3$ (as given by two opposite sides of a cube, centered on $\theta = \pi/2$). We do not make an effort to keep track of (statistical and systematic) errors in this appendix, as they are of secondary importance.

First we record the differential cross section for μ pair events:

$$\frac{d\sigma_{\mu\mu}}{d\Omega} = \frac{\alpha^2}{4s}(1+\cos^2\theta), \tag{C1}$$

where $\hbar = c = 1$, we have ignored the muon mass, and have set $\beta = 1$. Integrating over $4\pi$ gives the well-known total cross section (for the FQS beam energy)

$$\sigma_{\mu\mu} = \frac{4\pi\alpha^2(\hbar c)^2}{3s} = 103.3\,\text{pb}, \tag{C2}$$

where we have included the quantity $(\hbar c)^2 = 3.893 \times 10^8$ GeV$^2$ pb and used the FQS value of $s = 29^2$ GeV$^2$ to evaluate Eq. (C2). Using Eq. (C2) and the integrated luminosity $\mathcal{L}_{int} = 15.5$ pb$^{-1}$, we obtain $N(\mu\mu)_{tot} = \sim 1600$ events produced at the interaction point of the FQS experiment. In order to determine how many of these events will enter the $\Delta\Omega = 4\pi/3$ steradians of the FQS, we must integrate $\mathcal{L}_{int}$ times Eq. (C1) over the $\Delta\Omega$. That is, we write

$$\mathcal{L}_{int} \times 8 \int_0^{\pi/4} d\phi \int_0^{\xi_{max}(\phi)} \frac{\alpha^2}{4s}(1+\xi^2)d\xi = \sim 460\,\mu\text{ pairs}, \tag{C3}$$

where the limits on the angular integrations specify 1/8$^{th}$ of the $\Delta\Omega$ of the FQS, and $\xi = \cos\theta$. It is easy to show that for the FQS geometry, the $\xi$ limit,

$$\xi_{max}(\phi) = \frac{\cos\phi}{\sqrt{1+\cos^2\phi}}, \tag{C4}$$

is valid in the range $-\frac{\pi}{4} \leq \phi \leq \frac{\pi}{4}$.

To make the same estimate for the Bhabha events entering into the $\Delta\Omega$ of the FQS detector, we write

$$\mathcal{L}_{int} \times 4 \int_0^{\frac{\pi}{4}} d\phi \int_{-\xi_{max}(\phi)}^{\xi_{max}(\phi)} \frac{\alpha^2}{4s} \left(\frac{3+\xi^2}{1-\xi}\right)^2 d\xi = \sim 7600 \, \text{Bhabhas}, \tag{C5}$$

where only 1/4$^{th}$ of the detector $\Delta\Omega$ was covered by the integral because the distribution of Bhabha events, Ref. [26, p. 160], are not symmetrical in $\theta$ (but are skewed forward into the small $\theta$ region).

Using the results of Eq. (C3) and Eq. (22) of the text, it is easy to estimate as a function of $\beta$ (or, equivalently, magneticon mass) the number of magneticon pairs that one would expect to appear in the FQS apparatus. (In this region, for the 1$\gamma$ formulation, the relationship $\beta \Rightarrow Q$ holds to better than $10^{-3}$.) Thus, as an example, a magneticon of mass of 8.8 GeV/$c^2$, would locate a magneticon distribution (as calculated using the 1$\gamma$ formulation) just emerging from the low $Q$ tail of the main scatter plot distribution. This magneticon pair distribution would be comprised of ~230 magneticon pairs. For the 2$\gamma$ formulation, a magneticon of mass 12 GeV/$c^2$ would locate a magneticon distribution (of ~80 magneticon pairs) emerging from the low $Q$ tail of the main distribution. As mentioned in the text, assuming equal experimental spreads, the peaks of such distributions would be comparable to the line widths in the plots. Hence, they would probably not be noticed as a bump or a shoulder in the projected distributions – unless they were well clear of the main distribution.

[1] While the name "magneton" might seem appropriate for the set of magnetic spin ½ fermions contemplated in this paper, there would be confusion because the name magneton is already in common usage for other quantities, e. g., the Bohr magneton and the nuclear magneton. One might also consider using the name " magnetron," except that a magnetron is a well-known hardware RF source in common usage since WWII. It is also important to observe that the proposed family of magnetic particles are composite, deriving from the vorton model as described further in this paper and its references. Thus, the magneticon should not to be confused with any of the various highly charged monopoles, which have been discussed since Dirac's famous papers [1].) Therefore to choose a unique yet apt name, the word "magneticon" (for which there are no Google hits in 2014) was chosen.

[2] Ref. [2] introduced the word "dyality" to avoid possible confusion that could result from using the often used word "duality," in other contexts. (It is not surprising that as time passes, more applications for the concept of duality have been found.) As with the authors of Ref. [2], we also prefer to use the word "dyality" for this symmetry of generalized electromagnetism.

[3] This assumption is made since the lightest leptons are lighter than the lightest baryons.

[4] Cabibbo and Ferrari [4] eliminate the need for a second photon by imposing what they call "Zero Field Conditions" (ZFC). On the other hand, Wei and Baylis [5] have shown that the imposition of the ZFC sacrifices dyality invariance. To maintain dyality invariance, then, calls for a second photon, which one associates with a second, or magnetic, field tensor (or, equivalently a second, or magnetic, vector potential).

[5] The basis of our argument for the non-observation of a massless magnetic photon is an application of Bohr's correspondence principle. That is, in the generalized classical electrical electromagnetism of Ref. [6], electric charges and currents are the source for the (electric) vector potential $A$ and its associated (electrically sourced) fields, while magnetic charges and currents are the source for the magnetic vector potential $M$ and its set of magnetically sourced fields. Consequently, we argue that in a quantum mechanical description, electric charges radiate only electric photons and magnetic charges radiate only magnetic photons. It will be argued later in the paper that this restriction can be transferred into the Feynman diagrams of QED or rather QEMD. This argument, of course, entails some possible ambiguities and assumptions. Hence, experimental confirmation is called for.

[6] The distinction between the particles of the electric and magnetic sectors, as described in Ref. [3], is found in the electromagnetic modification in particle sub-structure at the most fundamental level, brought about by the application of the operator $\exp(\gamma_5 \Theta_d)$, where $\Theta_d = \pm \pi/2$.

[7] In addition, based upon this derivation, a quantization condition on electric and magnetic charges was obtained that entails no Dirac strings or other undesirable physical encumbrances.

[8] The idea to employ spacetime algebra (sometimes called Dirac algebra) to incorporate magnetic monopoles into classical electromagnetic theory was proposed by de Faria-Rosa *et al* . [13].

[9] We note that it is usual to omit the explicit notation, $\mathbf{1}_4$, and simply write this unit matrix as 1.

[10] This continuous symmetry of the homogeneous set of Maxwell's equations was first pointed out by Rainich [16].

[11] As a result of this (continuous) symmetry, Noether's theorem [17] predicts a conserved quantity, which we call dyangular momentum. Dyangular momentum plays a key role in the author's "A Model for Ball Lightning." [18]

[12] This classical interaction term finds an explicit analogue in quantum theory [75]. We use this analogy and Eq. (4) to furnish a guide to the appropriate signs, and placements of the $\gamma_5$ factors for the Lorentz force cross terms of QEMD.

[13] Blagojević and Senjanović [19] also use the acronym QEMD (for quantum electromegnetodynamics) but their formulation (as well as others that they reference) is based upon a one-potential model, with all of its associated difficulties (Dirac veto, loss of Lorentz invariance, potential singularities, etc.), rather than the two-potential model [6] that is the basis for this analysis.

[14] It has been customary to assume that the electromagnetic fields (*E* and *B*) generated by magnetic charges and currents would be of the same essence as those generated by electric charges and currents and, hence, would add directly to the usual electromagnetic fields, jointly forming a common $F_{\mu\nu}$. This requirement led to the conclusion that, while electric charge is a scalar quantity, magnetic charge must be a pseudoscalar quantity. In Ref. [6] electromagnetism and magnetoelectricity are unified by the concept of a common essence of generalized electromagnetic charge for both. Hence, both electric and magnetic charges are scalar quantities, interchanging places by the dyality rotation generated by the operator $\exp(\gamma_5 \Theta_d)$. This approach leads to the electromagnetic fields being of opposite parity to those of magnetoelectricity, and to distinct quantities $F_{\mu\nu}$ and $G_{\mu\nu}$, and to two physically distinct photons.

[15] Also, this explains why the $\gamma_5$ factor is present in the Lorentz force cross terms: the various terms in the Lagrangian must be Lorentz scalars.

[16] N. B., the (naïve) substitution $A_\mu \gamma^\mu \Rightarrow M_\mu \gamma^\mu \gamma_5$ was used in Ref. [20] to obtain Eq. (6) of this paper: Ref. [20, Eq. (12)]. This step introduces a $\gamma_5$ into the equation, converting a standard QED calculation into a proposed QEMD formulation. As will be seen, this step gives the same non-relativistic cross section as the more systematic relativistic QEMD evaluation of Fig. 1a found in Appendix A.

[17] See Ref. [20] for references. Also, see Ahlen's review paper [22]. In view of the estimated factor of ~two found for the cross section for the more general electron-magneticon scattering [see Eq. (A28)], this agreement may be fortuitous; there are no classical experimental results for e-g scattering for a proper comparison. It is evident that Eqs. (5 and 6) lead to a result that we would categorize as a one-photon formulation (in which category the earlier classical calculations would also fall), while Eq. (A28) is the result of what we are calling a two-photon formulation. That is, both the electric and the magnetic photons participate (it is a *t*-channel interaction) and contribute in roughly equal amounts. It was also shown in Appendix A that these two amplitudes don't interfere.

[18] We mention here that this magnetic Coulomb scattering interaction takes place in the *t*-channel because we suggest in this paper that there is an intrinsic difference between *t*-channel and *s*-channel processes. N. B., we argue below that in the *s*-channel magneticon pair production process, only one photon (the electric photon) participates.

[19] Using and extending the template of Ref. [21] via Eq. (4), then, the concept of the LFc term and the use of a $\gamma_5$ in QEMD are introduced.

[20] At this point we see that while the usual QED interaction is a vector interaction, with a $\gamma_\mu$ on the vertex, the LFc terms in this generalized EM theory are associated with an axial vector interaction, with a $\gamma_\mu \gamma_5$ (or a $\gamma_5 \gamma_\mu$) on the LFc vertex.

[21] To enable a more direct comparison to Eq. (8), we have set $Q = 1$, i. e., $Z = 1$ in recording the result in Ref. [21. Eq. (7.22)] – here written as Eq. (9).

[22] This selection for the value *e* leads to some sign differences with the equations of Ref. [21], in which *e* is defined as the electron charge.

[23] From a physical point of view, one could argue that such vacuum currents are fully shielded by the other particles residing in the (fully occupied) Dirac sea. A full-fledged mathematical argument can be found in papers using the front form description of dynamical systems (first introduced by Dirac [23] and further developed by numerous authors). In particular, cf. Refs. [24 and 25].

[24] We intend to study in a later paper this apparent violation of time invariance.

[25] As asserted by Dirac [23], the requirement of relativistic invariance does not appear to entail the requirement for invariance under time reflection.

[26] Actually, it may be appropriate to soften this statement somewhat. We shall argue in a later paper that a major component of dark matter may be in the form of magnetic hydrogen.

[27] They point out, of course, that in this instance they are considering the proton to be a structureless point-like particle, like the electron or muon.

[28] The vertex factor for the electron is put in with a plus sign because the quantity of charge *e* is defined in this paper (and in Ref. [26]) to be positive, while the electron itself is a negatively charged fermion. The sign of the charge of the heavy fermion, then, is carried by the indicated charge factor $eQ_f$, and the usual minus sign is attached to the vertex factor.

[29] This step away from well-established physics principles in some ways resembles the early days of the development of quantum mechanics. It was then important to experimentally verify the applicability of the new quantum equations with known classical results. Of course, the application in this case is somewhat more tenuous because the existence of the relevant classical results involving magnetic charge and currents has not yet been established.

[30] This vertex term carries a plus sign because the Clifford Lagrangian cross term $+\gamma_5 kA$ carries a plus sign.

[31] Of course, it is assumed that the usual generalized classical Lorentz force equation [6] (which includes magnetic charges and currents) is valid.

[32] Of course, it would be useful to experimentally explore whether the magnetic vacuum Maxwell production vertex is forbidden or just highly suppressed (partially enabled via vacuum fluctuations?).

[33] If, indeed, magneticons are experimentally produced, this suggestion about two distinct Feynman diagrams, that yield two distinct topologies, can be verified by looking at the ionization losses for magneticons traversing matter.

[34] The "doesn't matter" demonstration was actually shown in [21] for the case of electron-proton scattering.

[35] We use here the designation "configuration space" instead of "coordinate space" because for the purpose of integration the interaction points can be anywhere in space-time. That is, the amplitudes and cross sections are functions of the full field of space-time interaction points rather than any specific space-time point.

[36] As mentioned above, analysis using the front form or light-front QED [24, 25] give support to this point of view.

[37] Since these predicted magnetic particles, i. e. magneticons, have not yet been observed, it is clear that this proposed dyality symmetry, if it exists, is broken: i. e., the magneticon is much heavier than the electron (or muon, or even the tau).

[38] Further details about how Eq. (23) could come about will be discussed in a later publication.

[39] The notion that these two diagrams are topologically distinct is supported by the fact that their mathematical evaluations yield different results (although they are, of course, of the same order of magnitude).

[40] The factor of ~2, which is the result of the assumption that there are two topologically distinct Feynman diagrams and two distinct photons, is valid near the point of minimum ionization. It grows to a factor of ~4 at low values of ($\gamma - 1$), $10^{-4}$, say.

[41] As will be described, this effort has been carried out by some experimental groups, but they were guided by the idea, growing out of the analysis of Ref. [1], that magnetic monopoles would carry a large charge. It is not surprising, then, that the results of Ref. [1] led to a focus on large magnetic charge, well above the charge $g = e$. For success in a search for magneticons as described in this paper, an effort must be made to extend the experimental search down to a magnetic charge $g = e$, where, unfortunately, significant backgrounds of SM particles of unit charge would have to be addressed.

[42] In Appendix B we introduce the Calibri font $x$ to indicate the stopping power in MeV-cm$^2$/g, leaving $x$ to indicate the stopping power in ergs/cm. In general statements, this distinction is not necessary.

[43] The Crystal Ball Experiment at SPEAR at SLAC [33] had a maximum beam energy of ~3.7 GeV and, hence, falls within the limit already imposed by the DESY data [30]. Other $e^+e^-$ colliding beam experiments are listed in Ref. [34], but a perusal of these experiments reveals that they make no additional contributions to closing the excluded mass gap already described in the text.

[44] Of course, if a larger rate of magneticon pairs would be "unnoticed," then the upper limit of the magneticon mass exclusion range would be lower.

[45] The same apparatus as used for the exclusive free quark production search [28] was also used for an inclusive quark production search [37]. However, the inclusive quark search event trigger required three or more ionizing tracks, making it unsuitable for our present pair production study.

[46] Another technique to look for Dirac magnetic monopoles in cosmic rays is the use of Cherenkov radiation detectors, e. g., by the BAIKAL Collaboration [57]. However, these, too would not detect a magneticon of charge 1 $e$.

[47] To a first approximation this statement is true, but a number of corrections apply in a more detailed analysis [39]; cf., also Refs. [43, 44]. In addition, in the ultra-relativistic regime (not the case for this present work), nuclear scattering becomes significant.

[48] This is, of course, true because all projectile masses under consideration are much greater than that of the electron.

[49] It is unfortunate that Ref. [28] offers no explicit $\beta$ information for the events in their data set.

[50] Since at the time of the FQS experiment, there was speculation of possible fractionally charged leptons [40], the FQS experiment also used their data to rule out fractionally charged leptons for masses below 14 GeV/$c^2$.

[51] It appears that there are some typos (omission of the exponent 2) in the discussion found in Ref.[28]; a somewhat more detailed formulation [but the function f($\beta$) is undefined] is to be found in Ref. [37].

[52] The FQS has time of flight (ToF) counters, which were used to determine particle velocity, which in turn is used to estimate the expected stopping power. Since there is a $\beta^2$ in the denominator of the Bethe formula, the expected $dE/dx$ would be enhanced (in comparison to the minimum stopping power) by a $v^2$ factor electrical particles (of any charge). However, looking more closely at the Bethe $dE/dx$ formula for stopping power, one can see that the( reduction in) stopping power is not exactly proportionally to $\beta^{-2}$, but rather somewhat less (due to the log term). In the analyses of the data taken with the FQS apparatus [28, 37, 42] the effects of the log term were (evidently) taken into account with a phenomenological correction function.

[53] As we have noted in the text, at the center of the FQS blind spot in the 2γ formulation ($m_m = 10.7$ GeV/$c^2$), the ionization and $\beta$ (ToF measurements) of putative magneticons are consistent with those of an electric particle of unit charge. However, one would have to ask why would a pair produced electric particle would have a $\beta = 0.675$. As such a distribution would move away from this center point (with a change of the putative magneticon masses), the observed $\beta$ and $dE/dx$ would become less consistent with those of electrically charged particles. But the events in such a distribution, being only a few percent, would not show up distinctly in the scatter plot, and, hence, would be easy to overlook.

[54] We note that if a random collection of $N$ magneticons (of both polarities) were trapped in a sample, then the rms step size associated with a typical SQUID run for this sample would be proportional to $e_m\sqrt{N}$.

[55] The magnitude of a SQUID detector signal is independent of the monopole velocity as long as it passes all the way through the detector coil.

[56] The arguments of this section do not consider primordial cosmic monopoles, which are not expected to be associated with air showers

[57] It has recently been brought to our attention that a search for Drell-Yan pair produced Dirac monopoles has been carried out at the LHC by the ATLAS Collaboration [65]. No Dirac monopoles in the charge range $0.5g_D < g < 2.0g_D$ were detected in 7.0 fb$^{-1}$ of pp collision data. Unfortunately, in tailoring their trigger to detect highly ionizing tracks (which would be expected to characterize Dirac monopoles), magneticon tracks would be missed. (An earlier ATLAS paper is also available analyzing 2.0 fb$^{-1}$ of data [66]).

[58] It might be interesting to make an effort to mine data already taken at B factories (whose experimental lifetime is presently over), but such a data mining would entail a significant amount of effort for only a limited range of (possible) additional magneticon mass reach.

[59] We argue that the topological factor of $2^4 = 16$ should be only 15 ($= 16 - 1$); the basis for this argument is that a closed magneticon loop must carry at least one LFc term vertex to be called into existence in a QEMD interaction. The logic is analogous to that for vacuum bubbles, as discussed by Brodsky and Shrock [25], who use a light front analysis to show that vacuum bubbles cannot spontaneously form and hence do not contribute to the cosmological constant.

[60] The 2γ result derives from multiplying the 1γ result by the number of additional Feynman diagrams (15, see prior endnote) facilitated by the magnetic photon. Should the lightest magneticon be heavier than the 5.1 GeV/$c^2$ that was used in the calculations to evaluate the magneticon contribution to $a_e$, these contributions would be smaller.

[61] At times, we will also use the word magneticon to describe other spin ½ fermions.

[62] Actually Ref. [6] predicts four generations, but exploration of that prediction (assuming that magneticons are found to exist) would have to take place some time in the future.

[63] A template for such a result in the μ–e system has been published [76]; in Ref. [76], the Lagrangian has a permutation symmetry between muon and electron, but it is shown that there can be asymmetric solutions in which the physical masses differ. (This paper predates the Higgs papers and discovery.) Earlier references to the idea that there can be asymmetric solutions to a symmetric Lagrangian are also given in Ref. [76].

[64] This result follows from the assumption that both electric and magnetic charge are scalar quantities, as would be required by dyality symmetry.

[65] And, as we have noted above, using a version of Hamilton's principle, Ref. [6] presents a derivation of both the generalized Maxwell's equations *and* the generalized Lorentz equation from the same Lagrangian.

[66] QEMD stands for the QED in an electric world in which we try to account for magnetic charge and their LFc terms. Similarly, we use QMED to represent the physics in a (predominantly) magnetic world in which we introduce interactions involving electric charges and their associated LFc terms.

[67] While the assignment of vertex identities to photons connected to external fermion legs is (more or less) clear, e. g., pair production or particle scattering, the situation is more ambiguous for internal fermion loops.

[68] Ramsey [78] makes the argument that one needs to also include magnetic charge conjugation $M$ when considering the conjugation symmetries (the others being *TPC*) to require of acceptable physics equations and interactions. (One recalls that weak interactions were found to violate *C* conjugation, but not *CP* conjugation.)

[69] The specific details of this assertion will be explored in a later paper.

[70] We have already pointed out [68] that for magneticons of mass well into the multi GeV region the experiments at the LHC would be producing magneticons at the same rate as μ pairs (i. e., copiously).

[71] We argue that it is legitimate to move the $\gamma_5$ factor along a photon line which attaches to two different fermion lines because what the $\gamma_5$ factor does at a vertex is permute the EM interactions at the relevant vertex between the large and the small fermion components, and it doesn't make any difference which end of the internal photon line that this permutation would take place. We believe that the validity of this view depends upon the fact that a trace summation is applied to internal VP loops. Thus, in this case, the distinction between the Maxwell source vertex and the Lorentz force vertex is moot.

[72] Of course, for this to be a full identity, we depend upon the validity of Eq. (23).

[73] Knecht uses $\alpha = e^2/4\pi$.

[74] Of course, what we are doing here is making an estimate rather than a calculation. One would expect some differences (presumably minor) between our estimate and a proper calculation.

[75] Ahlen [22] gives an extensive review of the history of *dE/dx* for electrically charged particles. He also extends this to the stopping power for magnetically charged particles. However, since he used the one-photon formulation, we choose to use the Jackson Bethe formula as our reference starting point to develop our two-photon formulation. The Jackson Bethe formula is simpler in that it doesn't include all of the correction terms found in Ref. [41], but for the purposes of this paper, these corrections are not large enough to affect the basic conclusions of the analysis; to try to include them would add complications without adding any real contribution to the accuracy of the final results.

[76] The minimum *dE/dx* values given by NIST [73] for various materials agree well with those given by the PDG [38].

[77] In Ref. [22] (also, cf. Ref. [41]), Ahlen's Eq. (5.7) contains three (relatively) small correction terms: $-\delta_m/2$, $K(|g|)/2$, and $B(|g|)$, which we shall neglect. The density correction $-\delta_m/2$ is small and tends to vanish as the projectile kinetic energy $(\gamma - 1)m_p$ drops below the minimum ionization point; the $K(|g|)$ is an empirical correction obtained by Ahlen from a derivation by Kazama, et al. [74] for large magnetic charge; and $B(|g|)$ is a correction also associated with large magnetic charge, i. e., $|g| \sim 137e/2$ or $137e$, which we do not believe is relevant for our magneticon energy, mass, and charge.